\documentclass[11pt]{article}

\usepackage[margin=1in]{geometry}

\usepackage[numbers,sort&compress]{natbib}
\usepackage{authblk}

\usepackage[T1]{fontenc}
\usepackage{microtype}        %
\usepackage{graphicx}
\usepackage{booktabs}         %
\usepackage{tabularx}         %
\usepackage{array}
\usepackage{longtable}        %
\usepackage{listings}         %
\usepackage{xcolor}
\usepackage{tikz}
\usetikzlibrary{positioning, arrows.meta, calc}
\usepackage{amsmath,amssymb}
\usepackage[section]{placeins} %
\usepackage{hyperref}         %
\usepackage{cleveref}         %

\DeclareUrlCommand\fpath{\urlstyle{tt}}
\makeatletter
\g@addto@macro\UrlBreaks{\do\.\do\-}  %
\makeatother

\definecolor{yamlKey}{RGB}{0,90,156}
\definecolor{yamlStr}{RGB}{163,21,21}
\definecolor{yamlComment}{RGB}{0,128,0}
\definecolor{listingBg}{RGB}{248,248,248}

\lstdefinelanguage{yaml}{
  keywords={true,false,null},
  keywordstyle=\color{yamlKey}\bfseries,
  sensitive=true,
  comment=[l]{\#},
  commentstyle=\color{yamlComment}\itshape,
  stringstyle=\color{yamlStr},
  morestring=[b]',
  morestring=[b]",
  moredelim=[l][\color{yamlKey}]{-\ },
}

\newcommand{\ScientificEvidence}{\textsc{ScientificEvidence}}
\newcommand{\EvidenceVariable}{\textsc{EvidenceVariable}}
\newcommand{\EvidenceAssertion}{\textsc{EvidenceAssertion}}
\newcommand{\GeneticEvidence}{\textsc{GeneticEvidence}}
\newcommand{\GeneticEvidenceVariable}{\textsc{GeneticEvidenceVariable}}
\newcommand{\GeneticEvidenceAssertion}{\textsc{GeneticEvidenceAssertion}}

\newcommand{\dimval}[1]{\textsc{\lowercase{#1}}}


\begin{document}

    \title{A Semantic Model of Genetic Evidence: A Step Toward Bridging the Basic-Science--Clinic Gap}

    \author[1,2,3]{Michael Bouzinier\thanks{Correspondence: \texttt{michael\_bouzinier@harvard.edu}}}
    \author[3,4]{Dmitry Etin}
    \affil[1]{Harvard University, Cambridge, MA, USA}
    \affil[2]{IDEXX Laboratories, Westbrook, ME, USA}
    \affil[3]{Forome Association, Newton, MA, USA}
    \affil[4]{Deggendorf Institute of Technology, Germany}
    \date{}

    \maketitle

    \begin{abstract}
        Scientific and clinical decision-making depends on evidence from
        the primary literature, but existing standards for representing
        that evidence (FHIR Evidence, ECO, SEPIO, and the GA4GH Genomic
        Knowledge Standards) are oriented toward clinical-trial workflows,
        evidence codes, or single-variant assertions, and do not capture
        the fine-grained, domain-specific structure of claims in basic and
        pre-clinical research. We introduce a semantic model for scientific
        evidence with three core classes, specialize it for genetics, align it structurally to FHIR Evidence with a
        SEPIO-anchored credibility decomposition, and attach a compact
        dimensional vocabulary whose conditional-activation rules are
        validated by a SHACL schema for the implemented constraints. Using clinical variant
        interpretation as the driving use case, we evaluate the model
        through a human--AI annotation pilot over six genetics papers,
        yielding 28 evidence items and 95 source-anchored assertions, with
        a workflow that keeps curator-authored reference annotations distinct from
        AI-drafted annotations. Treating the pilot as a feasibility study
        rather than a benchmark, we argue that the model is a useful
        increment toward trustworthy, AI-ready infrastructure for variant
        interpretation: a reference data model and validation schema for
        representing genetic evidence.
    \end{abstract}


\section{Introduction}
\label{sec:intro}

Scientific and clinical decision-making depends on evidence from
heterogeneous sources, especially the primary literature. Today,
this evidence is largely consumed by humans and encoded indirectly
into annotation databases and local rules.

Existing resources such as HL7 Fast Healthcare Interoperability
Resources (FHIR) Evidence and EvidenceVariable~\cite{fhir_evidence},
the Evidence and Conclusion Ontology (ECO)~\cite{eco2019}, the
Scientific Evidence and Provenance Information Ontology
(SEPIO)~\cite{sepio}, and the Global Alliance for Genomics and
Health (GA4GH) Genomic Knowledge Standards~\cite{ga4gh_va}
advance interoperability in clinical research. However, they do not
comprehensively represent the evidence types and workflows of the
basic sciences. They focus mainly on clinical trials, high-level
assertions, and variant representations, with limited support for the
fine-grained, domain-specific evidence from basic and pre-clinical
research needed for automated reasoning and AI-ready infrastructure.

Each scientific domain demands a tailored approach for structuring
and communicating evidence. For example, the genetic evidence
discussed in this paper is generated in diverse contexts, from
experimental model systems to large-scale population studies, that
are not fully supported by current standards.
To properly use evidence derived from
scientific literature in clinical workflows, two principal challenges
require standardization: \emph{Relevance and Specificity}, that is,
how precisely a genetic finding maps to an individual patient,
disease context, or biological hypothesis; and \emph{Confidence},
that is, whether a given piece of evidence can be robustly and
safely used, ensuring defensible compliance, interpretability, and
reproducibility.

A well-designed schema for genetic evidence helps make literature
review a \emph{semantic-parsing} problem: mapping natural-language
claims to a shared, source-anchored representation. Large language
models are increasingly capable of this kind of parsing, but
dependable automation still requires a carefully structured target
language. The present work offers one such increment. Building on decades of work by
the biomedical-ontology community, it takes a small step toward literature
review that is materially assisted by automated parsing.

\paragraph{Contributions.} This paper makes the following contributions:
\begin{itemize}
    \item A domain-analysis-driven semantic model for genetic evidence, with
    core classes (\ScientificEvidence, \EvidenceVariable,
    \EvidenceAssertion), genetic specializations, a compact dimensional
    vocabulary, and a validation schema in the W3C Shapes Constraint
    Language (SHACL)~\cite{w3c-shacl}.
    \item An application of the model to six genetic-evidence papers spanning
        human, population, model-organism, and polygenic-score genetics,
        supported by extraction tooling.
    \footnote{\url{https://github.com/ForomePlatform/genetic-evidence-model}}
    \item A human--AI collaborative annotation workflow, packaged as executable
        AI skills for annotation and review that apply the model under curator
        supervision.
    \item A structural alignment of the core classes with HL7 FHIR Evidence,
        including a credibility representation that decomposes certainty into
        SEPIO-anchored facets. The dimensions are designed to align with ECO,
        SEPIO, IAO, and OBI; Supplementary Note~\ref*{supp:sec:supp-crosswalk} provides a partial
        term-level mapping.
    \item A curated term-level UMLS crosswalk of the dimensional vocabulary, in
        which every value is mapped with a typed relation or recorded as an
        argued gap. The crosswalk was produced with the open-source,
        model-agnostic \texttt{GEM Mapping Studio}, a standalone curation tool
        distributed as a Python package that supports the full workflow, from
        defining dimension axes in an empty workspace to adjudicating
        value-level mappings. The workflow follows a reproducible protocol with
        pre-registered search sweeps, structured rejection rationales, and a
        public decision log in the repository.
\end{itemize}

The remainder of the paper reviews related standards
(\Cref{sec:background}), defines the semantic model
(\Cref{sec:model}), and illustrates it with two worked examples
(\Cref{sec:examples}). It then describes the annotation workflow and
pilot statistics (\Cref{sec:workflow}), discusses findings, candidate
extensions, limitations, and future work (\Cref{sec:discussion}), lists
the released artifacts (\Cref{sec:availability}), and concludes
(\Cref{sec:conclusion}).

    \section{Background and Relation to Existing Standards}
\label{sec:background}

The representation of scientific and clinical evidence has been developed
across clinical informatics, biomedical ontology, and genomic knowledge
representation. Seven families of standards bear directly on our work: HL7 FHIR
Evidence~\cite{fhir_evidence}; the Evidence and Conclusion Ontology
(ECO)~\cite{eco2019} and the Scientific Evidence and Provenance Information
Ontology (SEPIO)~\cite{sepio}; the GA4GH Variant Annotation (VA) and Variation
Representation (VRS) specifications~\cite{ga4gh_va,ga4gh_vrs}; the ACMG/AMP
variant-interpretation guidelines~\cite{acmg2015}; the upper ontologies IAO
and OBI~\cite{iao2015,obi2016}; the PROV-O provenance model~\cite{prov_o};
and the ClinGen, MONDO, and HPO
curation and terminology resources~\cite{clingen2018,mondo,hpo}. Each captures
part of what a basic-science genetic publication reports, but none captures the
whole: a single conditional-knockout study may report molecular, cellular, and
clinical phenotypes at once, and a polygenic-score study may aggregate millions
of variants into one score. A per-family review, with the specific coverage
gaps that motivate our model,
is given in Supplementary Note~\ref*{supp:sec:supp-background}.
Supplementary Table~\ref*{supp:tab:coverage-matrix} presents the same
analysis as a matrix comparing ten evidence patterns across the standards
and our model (GEM). For each pattern, the matrix indicates whether a resource
represents it natively, only in part or by extension, or not at all. It
also marks patterns that lie outside a resource's subject matter.

\subsection{Relation to FHIR Evidence}
\label{sec:background-fhir}

FHIR is an HL7 information model rather than an OBO Foundry ontology. A
genetic-evidence model could, in principle, be built directly on the OBO
side, but doing so first would be premature. ECO and SEPIO provide,
respectively, an evidence-type vocabulary and a model of assertion
provenance and qualifiers.
Neither provides the variable-and-certainty content model for an evidence
unit that FHIR Evidence offers and that basic-science evidence needs to
support translation from bench to bedside.

Aligning a genetics-only model directly to OBO before such a general content
model exists would yield a narrow resource. It would neither compose cleanly
with other basic-science domains nor reuse OBO's orthogonal vocabularies,
creating exactly the redundancy that OBO's emphasis on orthogonality and
reuse is intended to prevent.

We therefore take the structural content model from FHIR, generalize it from
clinical trials to basic science, and design the dimensions to align with
ECO, SEPIO, IAO, and OBI. Supplementary Note~\ref*{supp:sec:supp-crosswalk} provides a term-level
crosswalk to these vocabularies, including verified identifiers and the
remaining gaps. Genetics is the first specialization; extending the model to
other basic-science domains and completing the term-level alignment are
future work.

The HL7 FHIR Evidence, EvidenceVariable, and EvidenceReport
resources~\cite{fhir_evidence} were designed with systematic-review and
guideline-development workflows in mind, where evidence concerns an
intervention tested in a population. FHIR Evidence therefore has a different native
framing and does not by itself represent the evidence produced by basic and
pre-clinical research. Our three core classes address this gap.

The core classes are general across the basic sciences rather than specific
to genetics. They retain FHIR's separation between the evidence variable and
the certainty of a finding, incorporate FHIR's separate \texttt{statistic}
into \EvidenceVariable{} as the measured value, and replace the
clinical-trial content frame of population, intervention, comparator, and
outcome with a genetics-specific dimensional frame. Genetics is the
specialization developed in this paper (\cref{sec:model}):
\ScientificEvidence{}, \EvidenceVariable{}, and \EvidenceAssertion{}
specialize to \GeneticEvidence{}, \GeneticEvidenceVariable{}, and
\GeneticEvidenceAssertion{}, respectively. \Cref{tab:fhir-alignment} gives
the class-level correspondence. The field-level alignment, including the
certainty subcomponents discussed in \cref{sec:model-dimensions}, is given in
Supplementary Table~\ref*{supp:tab:fhir-fields}.%
\footnote{We distinguish two notions that the literature often conflates.
The \texttt{assertion} property of a \ScientificEvidence{} item records the
object-level claim made by a paper. It is what the pilot records, and its
closest analogues are the GA4GH VA \texttt{Statement} and the SEPIO
\texttt{Assertion}, not the FHIR \texttt{Evidence.assertion} element, which
is a human-readable summary.

The \EvidenceAssertion{} class is retained for the reasoning layer developed
in~\cite{our_preprint}, where it is a meta-predicate over decision rules that
use such claims. That reasoning layer is not instantiated in this paper and
has no direct analogue in FHIR, VA, or SEPIO. In the released RDF
serialization, \texttt{GeneticEvidenceAssertion} currently types
object-level assertions so that assertion-level SHACL validation applies to
them. This temporary serialization choice does not assign the
meta-predicate role to the pilot assertions. The property and the class
share a name; a future schema revision will separate them.}

\begin{table}[tbp]
    \centering
    \caption{Class-level correspondence between the core model and FHIR Evidence
    R5. The alignment is architectural: the genetics-specific dimensional frame
    replaces FHIR's clinical-trial frame. Field-level detail is provided in
    Supplementary Table~\ref*{supp:tab:fhir-fields}.}
    \label{tab:fhir-alignment}
    \small
    \begin{tabularx}{\linewidth}{@{}llX@{}}
\toprule
This work & FHIR Evidence R5 & Relationship \\
\midrule
\ScientificEvidence
& \texttt{Evidence}
& Structural sibling; GEM replaces the PICO frame with a
  genetics-specific dimensional frame \\
\EvidenceVariable
& \texttt{EvidenceVariable} + \texttt{Evidence.statistic}
& Broader; GEM combines the variable definition and measured value in one class \\
\EvidenceAssertion
& no equivalent
& Meta-predicate over decision rules; part of the reasoning layer
  described in~\cite{our_preprint}, which is outside the scope of this paper \\
Credibility
& \texttt{Evidence.certainty}
& Structural parallel; both provide an overall rating with separable,
  genetics-specific subcomponents \\
\bottomrule
    \end{tabularx}
\end{table}

\section{A Semantic Model of Genetic Evidence}
\label{sec:model}

The model contains three parts: a small set of \emph{core classes}
(\cref{sec:model-classes}) that define what a unit of evidence is and
how evidence items compose; a \emph{dimensional vocabulary}
(\cref{sec:model-dimensions}) that describes each evidence item along
roughly twenty axes; and a \emph{conditional-activation} mechanism
(\cref{sec:model-conditional}) that keeps the vocabulary compact by
requiring dimensions only when they are semantically appropriate.

\subsection{Core Classes}
\label{sec:model-classes}

Three abstract classes form the core of the model:

\begin{description}
    \item[\ScientificEvidence] A unit of evidence, typically linked to a
          publication or dataset and described by a set of dimensions (see
          \cref{tab:dimensions-core,tab:dimensions-cond}). It carries an
          \texttt{assertion} property containing the object-level,
          source-anchored claims made by the source, for example that a
          variant is associated with a phenotype. These are the claims
          recorded in the pilot (\cref{sec:workflow}). A single paper may
          give rise to multiple \ScientificEvidence{} items when it reports
          distinct evidential claims, such as a case--control association and
          an \emph{in vivo} functional assay of the same variant.
    \item[\EvidenceVariable] A named quantity extracted from one or more
        \ScientificEvidence{} items, such as the gnomAD allele frequency of
        a variant, the LOD score in a family, or the odds ratio reported by
        a GWAS. Evidence variables are the terms from which predicates over
        Evidence are constructed. They are also described by dimensions,
        because the same quantity, such as an odds ratio, may carry different
        epistemic weight depending on how it was obtained.
    \item[\EvidenceAssertion] A meta-predicate, that is, a predicate about a
        predicate. It determines whether \EvidenceVariable{}s are used
        appropriately in a predicate over Evidence, namely in a decision
        rule. This class is the model's point of contact with the reasoning
        layer described in~\cite{our_preprint}, where such meta-predicates
        assert which evidence a rule may use. This paper constructs no decision rules and therefore instantiates no
        \EvidenceAssertion{} in its meta-predicate role. The class remains part of
        the model, but the interpretability and AI-readiness of the reasoning layer
        are the primary focus of~\cite{our_preprint}, not of the present pilot.
\end{description}

The \texttt{assertion} property and the \EvidenceAssertion{} class share
a name but denote different things. The property records the object-level
claims captured in the pilot, whereas the class is a meta-predicate over
any decision rule that would use those claims. The property name is
retained for compatibility with the released annotations and will be
renamed, for example to \texttt{selector}, in a future schema revision.

Specializing these classes for the genetic domain yields
\GeneticEvidence, \GeneticEvidenceVariable, and
\GeneticEvidenceAssertion. These specializations inherit the provenance
requirement and add the dimensional annotations described next.

\subsection{Dimensions}
\label{sec:model-dimensions}

Each \GeneticEvidence{} instance is described using a dimensional
vocabulary summarized in \cref{tab:dimensions-core}, which lists the
always-required dimensions, and \cref{tab:dimensions-cond}, which gives
representative conditional dimensions.\footnote{The full enumeration of
conditional dimensions, including those omitted from
\cref{tab:dimensions-cond} for space, is maintained in the project
repository at
\url{https://github.com/ForomePlatform/genetic-evidence-model/blob/master/schema/dimensions.md}.}

Six dimensions are \emph{always required}: Knowledge Domain, Method,
Target Type, Resolution, Credibility, and Phenotype Scale. No
well-formed genetic-evidence item is fully specified without them.
Together, they tell a downstream reader what kind of evidence the item
is, how it was obtained, what it concerns, and how credible it is in
context.

The dimensions differ in internal structure, and each is modeled
according to its own semantics rather than forced into a single
hierarchy. Knowledge Domain values are sibling framings rather than a
subsumption hierarchy. A single item may carry several values at once,
as when a functional study is interpreted in a clinical-genetics
context; the multi-valued cardinality captures this directly.

Phenotype Scale is an ordinal scale, from molecular to clinical, rather
than a class hierarchy. Modeling it through subsumption would therefore
misrepresent the dimension. Method, by contrast, is genuinely
hierarchical, and its \texttt{is\_a} relations are encoded in the SHACL
distribution. The chosen structure is therefore part of the definition
of each dimension.

Phenotype Scale and Variant Ascertainment were added during the annotation
work (\cref{sec:workflow}). Phenotype Scale distinguishes molecular
phenotypes, such as binding or activity, from cellular, histological,
organismal, and clinical phenotypes. It is independent of the scale of the
genetic target.

Variant Ascertainment records how a variant entered a study. It is meaningful
only for variant-level targets and is therefore a \emph{conditional}
dimension, required when Target Type is \dimval{Variant}. Both dimensions
are revisited in \cref{sec:discussion}.

Credibility records the methodological soundness of an asserted finding, and
nothing else. It is a structured decomposition rather than an open key-value
field: one overall ordinal rating together with separately rated facets for
statistical power or sample size, independent replication, multiple-testing
control, ancestry or population-stratification control, and ascertainment.
The facets remain separate rather than being collapsed into a single score
because clinical and research users may weigh them differently.

This separation is inspired by GRADE's general practice of assessing the
factors that affect certainty separately~\cite{grade2008}. GEM does not
adopt the GRADE framework or its rating rules, which were developed for
bodies of clinical evidence, particularly randomized and observational
studies. The GEM facets instead support credibility judgments about
individual evidence items from basic and pre-clinical research.

The overall rating is anchored on SEPIO confidence, defined as the degree of
belief that an asserted proposition is true~\cite{sepio}, and is implemented
as a genetics-specific value set bound to that attribute, as prescribed by
SEPIO profiles (Supplementary Table~\ref*{supp:tab:credibility-facets}).

Two constructs are excluded from credibility by design. First, the
direction or hedging of a finding belongs to the assertion itself and is
represented by the width of the value range it constrains. An honestly
reported null result from a rigorous study may therefore have high
credibility. Second, the extent to which evidence bears on a particular
question is applicability. It is captured by Relevance and Specificity and
computed for each question. Combining methodological soundness, assertion
content, and applicability in one field would conflate distinct constructs.

The remaining dimensions are \emph{conditional}: they are required, or
even meaningful, only when other dimensions take particular values. For
example, mode of inheritance is not meaningful for a polygenic score, and
knockout type is not meaningful for a human case--control study. Rather
than marking such dimensions as ``optional'' and relying on convention, the
model assigns explicit activation conditions to each, as described in
\cref{sec:model-conditional}. These conditions specify when a dimension
must be present; otherwise, it may be omitted. This makes the requirement
status transparent to both human curators and automated validators.

\begin{table}[tbp]
\centering
\caption{Always-required dimensions of the \GeneticEvidence{} model.
These apply to every evidence item regardless of context. Phenotype Scale
was added to this set during the annotation work
(\cref{sec:workflow}). KD abbreviates Knowledge Domain; ``obj.'' and
``subj.'' denote objective and subjective, respectively.}
\label{tab:dimensions-core}
\small
\begin{tabularx}{\linewidth}{@{}llllX@{}}
\toprule
Dimension              & Value type        & Cardinality & Objectivity & Example values \\
\midrule
Knowledge Domain       & categorical       & multiple & obj.       & \dimval{Human Genetics}, \dimval{Population Genetics}, \dimval{Model Organism}, \dimval{Gene Function} \\
Method                 & categorical       & multiple & obj.       & \dimval{Statistical Genetics}, \dimval{In Vivo}, \dimval{In Vitro}, \dimval{Clinical Evidence} \\
Target Type            & categorical       & single   & obj.       & \dimval{Gene}, \dimval{Variant}, \dimval{Interval}, \dimval{Transcript} \\
Resolution             & categorical       & single   & obj.       & \dimval{Window}, \dimval{Gene}, \dimval{Functional Element}, \dimval{Position}, \dimval{Variant} \\
Credibility            & rating ${+}$ facets & multiple & subj.      & overall rating ${+}$ \texttt{sample\_size}, \texttt{replication}, \texttt{multiple\_testing}, \texttt{stratification\_control}, \ldots \\
Phenotype Scale        & categorical       & single   & obj.       & \dimval{Molecular}, \dimval{Cellular}, \dimval{Histological}, \dimval{Organismal}, \dimval{Clinical} \\
\bottomrule
\end{tabularx}
\end{table}

\begin{table}[tbp]
\centering
\caption{Conditional dimensions of the \GeneticEvidence{} model.
Each is required only when its activation condition holds. KD, MT, and TT
abbreviate Knowledge Domain, Method, and Target Type, respectively;
$\supseteq$ denotes set membership for multi-valued dimensions.
Dimensions marked $^{\ddagger}$ have their activation condition enforced by
an executable SHACL shape; for the others the condition is documented in
the schema and checked at curation time, with executable shapes scoped as
future work.}
\label{tab:dimensions-cond}
\small
\begin{tabularx}{\linewidth}{@{}lllX@{}}
\toprule
Dimension              & Value type   & Cardinality & Activation condition \\
\midrule
Mode of Inheritance$^{\ddagger}$ & categorical  & single   & KD $\supseteq$ \dimval{Human Genetics} AND TT = \dimval{Gene} \\
Mendelian Segregation  & boolean      & single   & KD $\supseteq$ \dimval{Human Genetics} AND TT = \dimval{Gene} \\
Penetrance             & categorical  & single   & KD $\supseteq$ \dimval{Human Genetics} AND TT = \dimval{Variant} \\
Variant Ascertainment$^{\ddagger}$ & categorical  & multiple & TT = \dimval{Variant} \\
Measurement Target     & categorical  & multiple & KD $\supseteq$ \dimval{Gene Function} \\
Gene Relation          & categorical  & single   & KD $\supseteq$ \dimval{Gene Function} \\
Organism$^{\ddagger}$ & categorical  & multiple & MT $\supseteq$ \dimval{In Vivo} OR KD $\supseteq$ \dimval{Model Organism} \\
Knockout Type          & categorical  & single   & KD $\supseteq$ \dimval{Model Organism} \\
\bottomrule
\end{tabularx}
\end{table}

\subsection{Conditional Activation}
\label{sec:model-conditional}

Conditional activation is the least conventional part of the model, but it
keeps the vocabulary compact without sacrificing expressive power. Rather
than using a fixed schema in which every \GeneticEvidence{} item has every
dimension, with many dimensions marked as optional and left empty, the model
requires particular dimensions only when specified conditions over other
dimensions hold. This mirrors expert reasoning. A curator reading a paper
about an X-linked recessive pedigree asks about mode of inheritance because
the evidence makes that question meaningful, not because a global schema
requires it.

Three activation conditions are currently encoded as SHACL shapes with
SPARQL-based constraints: Variant Ascertainment, Mode of Inheritance, and
Organism (marked in \cref{tab:dimensions-cond}). SPARQL-based constraints
are the standard SHACL mechanism for expressing requirements that cannot be
captured by simple cardinality constraints on a single property. Two
representative shapes illustrate the pattern. For brevity, we show only the
\texttt{sh:sparql} body of each and omit the outer NodeShape wrapper. The
full shapes, together with the base \texttt{GeneticEvidenceShape} and the
simpler \texttt{variant\_ascertainment} rule, are available in the project
repository with an annotated walkthrough. The remaining conditional
dimensions are documented in the schema and
assessed during curation;
executable shapes for them are future work.\footnote{\url{https://github.com/ForomePlatform/genetic-evidence-model/blob/master/schema/examples.md}}

Checking activation conditions is a closed-world question: whether an
annotation contains every dimension required by the evidence item. SHACL can
answer this question directly. OWL's open-world semantics cannot express
this form of completeness, which is one reason the constraints are encoded
in SHACL rather than as ontology axioms.

The first example encodes a conjunction across two classification axes.
\emph{Mode of Inheritance} is required when Knowledge Domain contains
\dimval{Human Genetics} \textbf{and} Target Type is \dimval{Gene}
(\cref{lst:shacl-moi}).

\begin{lstlisting}[language=yaml, caption={Compact form of the conditional-activation shape for \emph{Mode of Inheritance}. The outer NodeShape wrapper is omitted. The two triple patterns in \texttt{WHERE} form the conjunction; \texttt{FILTER NOT EXISTS} identifies instances that satisfy the condition but lack the required dimension.}, label={lst:shacl-moi}]
sh:sparql [ sh:select """
  SELECT $this WHERE {
    $this gem:knowledgeDomain gem:HUMAN_GENETICS .
    $this gem:targetType      gem:GENE .
    FILTER NOT EXISTS { $this gem:modeOfInheritance ?m }
  }""" ] .
\end{lstlisting}

The second example encodes a disjunction across two classification axes.
\emph{Organism} is required when Method contains \dimval{In Vivo}
\textbf{or} Knowledge Domain contains \dimval{Model Organism}
(\cref{lst:shacl-org}). The two disjuncts refer to different dimensions:
one is a Method value and the other is a Knowledge Domain value. This
illustrates that activation conditions can span the dimensional vocabulary
rather than being confined to a single axis.

\begin{lstlisting}[language=yaml, caption={Compact form of the conditional-activation shape for \emph{Organism}. The disjunction is expressed by a SPARQL \texttt{UNION} whose two branches refer to different dimensions.}, label={lst:shacl-org}]
sh:sparql [ sh:select """
  SELECT $this WHERE {
    { $this gem:method          gem:IN_VIVO . }
    UNION
    { $this gem:knowledgeDomain gem:MODEL_ORGANISM . }
    FILTER NOT EXISTS { $this gem:organism ?o }
  }""" ] .
\end{lstlisting}

In each case, the \texttt{SELECT} returns every \GeneticEvidence{}
instance that satisfies the activation condition but lacks the required
dimension. The validator reports these instances as constraint violations
with a localized diagnostic. Writing each conditional rule explicitly adds
some complexity, but it improves downstream clarity: a reader of an
annotation file can see why a particular dimension is required or absent,
and a validator can check conformance automatically.

\Cref{sec:workflow} returns to conditional activation in the context of
the annotation workflow.

\section{Illustrative Examples from the Literature}
\label{sec:examples}

To demonstrate the model's coverage and identify where it reaches its
current limits, we applied it to six published genetic-evidence papers that
span the major epistemic shapes encountered in literature-based variant
interpretation (Supplementary Table~\ref*{supp:tab:corpus}). The set covers
classical Mendelian human genetics (Gupta), population-scale association
(Duerr), case--control resequencing (Davis), model-organism and \emph{in
vitro} functional genetics (Jossin, Nelson, and the functional components of
Davis), and a genome-wide polygenic score (Inouye). Four of the six papers
were annotated manually from curator PDF highlights and callouts and form
the curator-authored reference set for this feasibility corpus. The remaining two were drafted by an AI assistant
and reviewed by a curator. All annotations and per-paper case reports are
available in the project
repository.\footnote{\url{https://github.com/ForomePlatform/genetic-evidence-model/tree/master/case-reports}}

Duerr et al.\ 2006 (\cref{sec:ex-duerr}) is the reference example of a
classical variant-level association study. Inouye et al.\ 2018
(\cref{sec:ex-inouye}) is a polygenic risk score study that exposes
limitations of the current variant-level model and motivates a cluster of
candidate extensions (\cref{sec:discussion}).

\subsection{A Genome-Wide Association Study Example}
\label{sec:ex-duerr}

Duerr et al.\ 2006 identified \textit{IL23R} as an inflammatory bowel
disease gene through a genome-wide association study of non-Jewish patients
with ileal Crohn's disease, followed by replication in a Jewish cohort and
family-based transmission testing. Under our decomposition, the paper yields
seven \GeneticEvidence{} items (Supplementary
Table~\ref*{supp:tab:corpus} and Case Report CR-DU): variant-level discovery
and replication, a family-based transmission analysis, conditional
fine-mapping, negative results at related pathway genes, and cited
functional and mouse-model support.

Six of the seven items fit the current schema without strain. The seventh, a
cited anti-p40 antibody trial, concerns drug-response evidence for which the
model has no fitting genetic Knowledge Domain, revealing a
translational-scope gap. Review of this annotation also identified three
candidate extensions that remain provisional under the two-papers rule: an
explicit direction-of-effect dimension for protective and risk-conferring
alleles, a construct for cross-item therapeutic synthesis, and an extension
of conditional activation from gene-level to variant-level Mendelian context.
Further detail is given in Supplementary
Note~\ref*{supp:sec:supp-ex-duerr}.

\subsection{A Polygenic Risk Score Example}
\label{sec:ex-inouye}

Inouye et al.\ 2018 developed and externally validated a meta-genomic risk
score (metaGRS) for coronary artery disease by combining three component
scores and evaluating performance in UK Biobank participants. Unlike Duerr's
variant-level association study, the evidential target is a derived,
whole-genome score and the primary measurements are epidemiological.

Our annotation decomposes the paper into four \GeneticEvidence{} items:
predictive and stratifying performance; independence from conventional risk
factors; persistence of stratification among medicated individuals; and the
headline finding that combining three component scores outperforms each
component (Case Report CR-IN). All four require deliberate placeholder
assignments in the current model. In particular, the score is assigned the
placeholder Target Type \dimval{Variant}, and Variant Ascertainment is marked
not applicable. These forced fits are recorded as six candidate extensions:
a score target type, whole-genome-aggregate resolution, a target-composition
axis, epidemiological measurement targets, a structured cohort descriptor,
and a flag for derived-artifact targets. The model's limitations are thus
recorded explicitly rather than left implicit. Further detail is given in
Supplementary Note~\ref*{supp:sec:supp-ex-inouye}.

\section{Human--AI Annotation Workflow and Evaluation}
\label{sec:workflow}

We designed and piloted an annotation workflow in which every annotation is
\emph{auditable}. Each assertion is anchored to the specific text that
supports it; each dimension value is defensible under the schema's
activation conditions; and each judgment that a downstream user might
contest carries an explicit reviewer or AI flag.

The same commitment applies to AI assistance. When an AI system drafts an
annotation, its uncertainties and assumptions are recorded for curator
review rather than silently absorbed into the annotation.
\Cref{sec:workflow-flags} reports the practical effect of that review. The
manual and AI-drafted tracks have different epistemic roles, primary and
secondary knowledge work, respectively, and use distinct flag vocabularies
to record provenance.

We treat the pilot as a \emph{feasibility study}, not as a benchmark.
Six papers and a single annotator do not support claims about
inter-annotator agreement or generalizability. The pilot shows that the
workflow's epistemic commitments can be implemented in practice and that
the schema exposes, rather than hides, its points of strain.

\begin{figure}[tbp]
  \centering
  \begin{tikzpicture}[
      box/.style={draw, rounded corners, align=center, inner sep=3pt,
                  text width=25mm, minimum height=9mm, font=\footnotesize},
      arr/.style={-{Latex[length=2mm]}, semithick}]
    \node[box, fill=black!5] (src) {Source publication (PDF)};
    \node[box, fill=blue!6, below left=12mm and 22mm of src] (m1) {Curator highlights \& callouts};
    \node[box, fill=blue!6, below=6mm of m1] (m2) {Structured JSON record};
    \node[box, fill=blue!6, below=6mm of m2] (m3) {YAML annotation\\(curator reference set)};
    \node[box, fill=orange!8, below right=12mm and 22mm of src] (a1) {AI draft\\(annotation skill)};
    \node[box, fill=orange!8, below=6mm of a1] (a2) {Curator review\\(review skill)};
    \node[box, fill=orange!8, below=6mm of a2] (a3) {Reviewed YAML annotation};
    \node[box, fill=black!5] (val) at ($(m3)!0.5!(a3) - (0,1.3cm)$) {SHACL validation\\(conditional activation)};
    \draw[arr] (src) -- (m1);
    \draw[arr] (src) -- (a1);
    \draw[arr] (m1) -- (m2);
    \draw[arr] (m2) -- (m3);
    \draw[arr] (a1) -- (a2);
    \draw[arr] (a2) -- (a3);
    \draw[arr] (m3) -- (val);
    \draw[arr] (a3) -- (val);
  \end{tikzpicture}
  \caption{The two annotation tracks. Curator-authored annotations
  (left) form the feasibility corpus's reference set. AI-drafted annotations (right) are
  produced under an annotation skill from the source PDF, the schema, and
  the manual annotations as exemplars, then reviewed by a curator under a
  review skill. Both tracks are validated against the same SHACL shapes.
  In every case, the curator, not the model, is the authority for the
  recorded annotation.}
  \label{fig:workflow}
\end{figure}
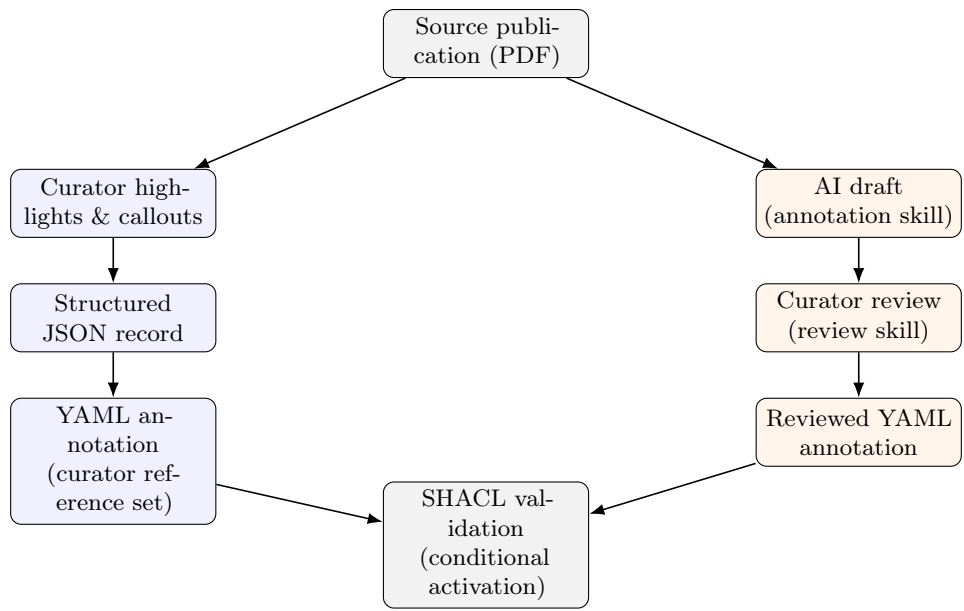

\subsection{Protocol and Coverage}
\label{sec:workflow-protocol}

Annotations are created through two tracks (\cref{fig:workflow}).

\paragraph{Manual track.}
Four of the six publications, Jossin, Davis, Nelson, and Gupta, were
annotated manually by Vladimir Seplyarskiy, a researcher in human and
population genetics. The curator read each paper, marked relevant
passages in the PDF with highlights and callouts, exported those
annotations with a helper script to a structured JSON record, and mapped
the result, together with a per-paper summary row, to a YAML annotation
conforming to the schema.

When this conversion required normalization, the change was recorded
inline as a \texttt{normalization\_note} with
\texttt{type = ai\_normalization}. For example, the free-text
``localisation and protein expression'' was mapped to the enumeration
values \dimval{Localization} and \dimval{Existence}. This record allows
later audits to compare the original and normalized values. The
manual-track protocol is described in more detail in Supplementary
Note~\ref*{supp:sec:supp-protocol-suite}.

\paragraph{AI-drafted track.}
For two publications, Duerr and Inouye, an AI assistant drafted the
annotations directly from the PDFs under an executable annotation
skill~\cite{anthropic_skills}. The assistant received the schema
(\texttt{dimensions.md}), the curator-authored reference annotations for the four
manual papers as exemplars, and the source PDFs. It produced YAML
annotations in the same format as the manual annotations.

When the assistant judged a mapping to be uncertain, or concluded that a
judgment call required human review, it recorded the issue inline as
either \texttt{ai\_uncertainty} or \texttt{ai\_assumption}. One of the
authors then reviewed the AI-drafted annotations assertion by assertion.
The reviewer has expertise in genomic data curation and the evidence
model, rather than in genetics domain science. The review assessed source
fidelity and schema conformance, not the underlying science. AI-generated
drafts and AI-assisted review suggestions were subject to final adjudication
by the human curator; the reviewed annotations are human-approved records,
and no AI system was treated as the authority for any scientific or
curation decision.
The four manual annotations are treated as the curator-authored reference
annotations for their own papers and as exemplars for AI drafting. The executable skills
and review protocol are described in Supplementary
Notes~\ref*{supp:sec:supp-protocol-suite}
and~\ref*{supp:sec:supp-skills}.

\paragraph{Source anchoring.}
In both tracks, every assertion, that is, the \texttt{assertion} property
of a \GeneticEvidence{} item, carries a \texttt{source\_span}. The span
identifies a specific page and includes a short key phrase, typically fewer
than fifteen words, from the source text. The phrase is long enough for a
reader to locate the passage by search while remaining short enough to
respect fair-use norms for copyrighted material.

Source anchoring is mandatory: an annotation with any assertion lacking a
source span is rejected at validation time. Across the six-paper pilot, all
95 assertions are source-anchored. This is a property of the workflow's
validation requirement, not an observed outcome.

\paragraph{Coverage.}
Dimension coverage across the 28 \GeneticEvidence{} items is summarized in
Supplementary Note~\ref*{supp:sec:supp-coverage-note}
(Table~\ref*{supp:tab:supp-coverage}) and in the repository's coverage
report.\footnote{\url{https://github.com/ForomePlatform/genetic-evidence-model/blob/master/annotations/coverage.md}}

\subsection{Flags and Extension Promotion}
\label{sec:workflow-flags}
\label{sec:workflow-promotion}

The two tracks use small flag vocabularies to make their epistemic
provenance explicit.

\paragraph{Manual track.}
The manual track uses three reviewer flags. A
\texttt{reviewer\_disagreement} records a clear factual correction, such
as the organism-set correction in Davis~2011. A
\texttt{reviewer\_suggestion} records an optional addition whose inclusion
depends on curator preference, such as HPO terms for patient phenotypes in
Davis, Nelson, and Gupta. A \texttt{reviewer\_query} records a case in
which the mapping to the schema is unclear and should not be silently
corrected. For example, the Nelson summary assigned
\texttt{specificity\_of\_phenotype = SNV}, which does not fit the
dimension's semantics.

\paragraph{AI-drafted track.}
Because there is no prior reference annotation with which to disagree, the
AI-drafted track uses two flags. An \texttt{ai\_uncertainty} records a
mapping that the assistant judged unreliable and referred for a human
decision, such as a borderline assignment of Mode of Inheritance. An
\texttt{ai\_assumption} records a judgment made without strong textual
support, such as whether cited mouse-model evidence should be represented
as a separate \GeneticEvidence{} item or treated as background context.

Across the six-paper pilot, manual-track annotations carried five queries,
six suggestions, and one disagreement, whereas AI-drafted annotations
carried six uncertainties and six assumptions. Of the nine AI-drafted items
from Duerr and Inouye, none was rejected: one was approved as drafted,
three were approved with non-blocking flags, and five were edited. The
later Inouye draft required fewer edits than the earlier Duerr draft, but
the papers differ in genre and protocol version; this comparison is
therefore suggestive rather than controlled. The edits mainly affected
dimension assignment and credibility calibration rather than the factual
content of assertions. The reviewer also added a small number of
higher-level items, such as a paper's overall thesis or a cited
translational study. Per-paper counts and detailed rationales are given in
Supplementary Note~\ref*{supp:sec:supp-ai-review}.

\paragraph{Extension promotion.}
The workflow adopts a conservative rule for evolving the schema. A
representational gap observed in one paper is recorded only in that
paper's \texttt{candidate\_extensions} block. A gap observed
independently in a second paper is promoted to a first-class dimension or
enumeration value.

In the six-paper pilot, this rule promoted two extensions. \emph{Phenotype
Scale} was first proposed in Jossin and then reused in Davis, Nelson, and
later papers. \emph{Variant Ascertainment} was first proposed in Davis and
then reused in Nelson and Duerr. Both now appear in the vocabulary
(\cref{tab:dimensions-core,tab:dimensions-cond}): Phenotype Scale among
the always-required dimensions and Variant Ascertainment among the
conditional dimensions, where it is required when Target Type is
\dimval{Variant}.

Fifteen other candidates remain unpromoted, pending a second independent
use or an explicit curator decision. The rule also prevents premature
additions. For example, a proposed polarity dimension for negative
assertions in Duerr was withdrawn when review confirmed that the existing
assertion predicate already represents both absence and presence.

\section{Discussion}
\label{sec:discussion}

Two features of the proposal merit brief emphasis. First, the
always-required vocabulary is small, and conditional activation keeps
the extended vocabulary compact without sacrificing expressive power.
Second, every assertion is anchored to a specific source span in its
source publication, making each annotation independently auditable.

\subsection{Positioning and Model Extensions}
\label{sec:disc-positioning}
\label{sec:disc-extensions}

The released schema formalizes the implemented conformance
requirements as SHACL shapes, makes the implemented activation
conditions machine-checkable, and anchors each assertion to its
source text rather than only to a style guide. The flag vocabularies
of \cref{sec:workflow-flags} make provenance explicit: a reader can
tell whether a value was drawn from the author's summary, proposed
by a reviewer, added as an AI assumption, or left deliberately
unpopulated. More broadly, the schema is intended to make literature review tractable as
a \emph{semantic-parsing} problem: the mapping of natural-language claims
to a shared, source-anchored representation. This is not a task a language model can perform in isolation. The target of
each parse is a structured, source-anchored representation, so the quality
of the parse depends on the quality of that target language. We see the present
contribution as one increment along a longer path, building on
decades of biomedical-ontology work (ECO, SEPIO, HPO, IAO, OBI, VRS,
and others), not as a replacement for any of it. A complementary
downstream layer, in which an epistemological type system and
machine-checkable meta-predicates constrain the decision logic that
consumes such evidence, is addressed in~\cite{our_preprint}.

The pilot identified two candidate dimensions that were exercised by
multiple papers and promoted to first-class status, Phenotype Scale and
Variant Ascertainment. It also identified fifteen further candidate
extensions that remain unpromoted under the two-papers rule
(\cref{sec:workflow-promotion}). The pattern of
candidates is summarized in \cref{tab:candidates}; per-candidate
rationale, affected dimensions, and proposed enumeration values
are documented in a companion \texttt{schema/EXTENSIONS.md} in the
repository.\footnote{\url{https://github.com/ForomePlatform/genetic-evidence-model/blob/master/schema/EXTENSIONS.md}}
Supplementary Table~\ref*{supp:tab:coverage-matrix} places these candidates
in a broader context. Of the ten evidence patterns considered, five common
patterns in basic-science genetic publications are represented natively by
the model. Three patterns were exposed by the pilot as forced fits and define
documented areas for extension. Two further patterns, therapeutic-response
evidence and cross-publication synthesis, are outside the model's intended
scope and are left to standards designed to represent them.
The Gupta annotation, which identified no candidate extensions, is also a
positive finding. It reports a family with an X-linked splice-site mutation
and careful pedigree segregation, a canonical evidence shape for which the
model was designed. The schema accommodates this evidence without strain.

\begin{table}[tbp]
\centering
\caption{Unpromoted candidate schema extensions surfaced by the
six-paper pilot, grouped by concern. Per-candidate detail is in the
companion \texttt{schema/EXTENSIONS.md}; paper keys are J~=~Jossin,
D~=~Davis, N~=~Nelson, DU~=~Duerr, IN~=~Inouye. Gupta surfaced none.}
\label{tab:candidates}
\small
\begin{tabularx}{\linewidth}{@{}llX@{}}
\toprule
Group & Papers & Count and examples \\
\midrule
Finer structural resolution  & J, N   & 4 (subdomain, interaction, gene-relation, cross-level variant) \\
Polygenic-score machinery    & IN     & 6 (score target, whole-genome aggregate, target composition, epidemiological measurement, cohort sub-block, derived-artifact flag) \\
Effect-size and curation     & DU, D  & 3 (direction of effect, knowledge-domain priority, curator critique) \\
Cross-item and activation scope & DU & 2 (cross-item/translational synthesis, conditional-activation scope) \\
\bottomrule
\end{tabularx}
\end{table}

\subsection{Limitations and Future Work}
\label{sec:disc-limitations}
\label{sec:disc-future}

The pilot has significant limitations. Six papers and a single
annotator do not support claims about inter-annotator agreement or
generalizability; we treat the work as a feasibility study, not a
benchmark. The evaluation is not a clinical
deployment: although the model is motivated by variant
interpretation in clinical-genomics programs, the pilot stops short
of applying the model to a live workflow.

Three near-term directions follow from these limitations. First, annotate a
second corpus selected to exercise the strongest candidate extensions.
Second, generate OWL, SHACL, JSON Schema, and LinkML renderings from a
single source of truth. This would open the schema to the OBO and Bridge2AI
ecosystems and support term-level alignment of the dimension vocabularies
with relevant ontologies, for example by aligning the molecular and cellular
phenotype scales with Gene Ontology biological processes. Third, conduct a
clinical-deployment study to test whether the feasibility claim translates
into utility.

A further limitation revealed by the UMLS crosswalk is representational
rather than empirical. Several GEM tokens are intersective compounds:
\dimval{statistical\_genetics} denotes statistics~$\times$~genetics
(\texttt{C0038215}~$\times$~\texttt{C0017398}), and
\dimval{bioinformatics\_inference} denotes inference~$\times$~bioinformatics.
Relational tokens have the same structure. For example,
\dimval{related\_gene} combines a relation operator with \emph{Genes}
(\texttt{C0017337}). Because the current crosswalk maps each GEM token to
one UMLS concept, it must sometimes use the nearest available anchor rather
than a concept that expresses the full compound meaning. For
\dimval{statistical\_genetics}, that anchor is \emph{Biostatistics},
recorded with the relation \emph{narrower}.

Concept-level post-coordination would address this limitation by mapping a
token to a small set of concepts whose intersection represents its intended
meaning. The UMLS Metathesaurus does not provide such a mechanism at the
aggregate level. As an aggregator, it records the concepts asserted by its
source vocabularies and the links among them, but it does not define a
compositional grammar of its own. Composition remains a feature of individual
source vocabularies, with the compositional grammar of SNOMED~CT being the
best-known example. It cannot be transferred directly across the aggregate:
the same intended intersection may be pre-coordinated in one source, absent
from another, and decomposed differently in a third~\cite{umls2004}.

The problem is particularly clear for \dimval{related\_gene}. UMLS often
contains pre-coordinated specific instances while lacking a generic concept:
gene-relatedness appears in terms such as \emph{modifier},
\emph{homologous genes}, and \emph{linked genes}, whereas the generic sense
of ``related in some way'' is represented only by the bare operator
\emph{Associated with} (\texttt{C0332281}). Thus
\dimval{related\_gene} can be represented as
\texttt{C0332281}~$\times$~\texttt{C0017337}, but not by a single concept.
Extending the crosswalk with an explicit composite relation, represented as
a structured set of concepts, is future work. At present, the intended composition
is preserved in the mapping's curator note; where the relation is the
semantic core, the operator concept itself is recorded as the anchor with
relation \emph{related}.

The crosswalk review also highlighted a representational limitation of the
Resolution axis. Its current enumeration presupposes genomic coordinates, but
genetic evidence may be localized in several coordinate systems. A variant
may be described at the genomic level (HGVS~\texttt{g.}), the transcript
level (\texttt{c.}), or the protein level (\texttt{p.}).

For example, an assertion of the form ``variant $V$ has effect $E$ in
transcript $T$'' is resolved in transcript coordinates. Such assertions are
often tissue-restricted and are characteristic of bioinformatics inference.
The domain mapping underlying the \dimval{PROTEIN\_SUBDOMAIN} candidate is
instead resolved in protein coordinates. In addition, \dimval{VARIANT}
itself has two components: a position and a discrete nucleotide change.

We therefore record \dimval{VARIANT\_IN\_TRANSCRIPT} as a curator-surfaced
candidate value for Resolution. It is separate from the fifteen
pilot-surfaced extensions listed in \cref{tab:candidates} and awaits
independent use in a corpus paper under the promotion rule.

\section{Availability of Code, Data, and Artifacts}
\label{sec:availability}

All artifacts referenced in this paper are open-source and
available in the project repository at
\url{https://github.com/ForomePlatform/genetic-evidence-model}.
This includes the SHACL schema, the human-readable dimension
reference (\texttt{dimensions.md}), per-paper annotations for all
six pilot publications, case reports, the companion
\texttt{schema/EXTENSIONS.md}, the SHACL walkthrough
(\texttt{schema/examples.md}), the annotation and review protocol
suite (\texttt{protocols/}) and the two executable annotation and
review skills (\texttt{skills/}) described in Supplementary Note~\ref*{supp:sec:supp-skills},
and the extraction tooling used to
convert PDF highlights and callouts into annotation scaffolds. The
namespace \url{https://w3id.org/genetic-evidence-model/} resolves
to the same repository. This paper describes release \texttt{v0.2.5},
archived on Zenodo under the concept DOI
\url{https://doi.org/10.5281/zenodo.22260686}, which resolves to the
latest archived release;
the GEM Mapping Studio is distributed on PyPI as
\texttt{gem-mapping-studio}.

\subsection*{Reproducibility and conformance}

All counts reported in this paper were generated from the accompanying
tagged release. From a clean checkout, corpus validation and regeneration of
the coverage table require two commands: \texttt{gem-validate} and
\texttt{gem-coverage}. Continuous integration runs both checks on every
change.

The paper distinguishes SHACL-enforced activation conditions, marked in
\cref{tab:dimensions-cond}, from documented rules that are not yet
executable. The enforced activation shapes are unit-tested in both
directions: a conforming instance passes, whereas an instance that satisfies
an activation condition but lacks the required dimension is reported as a
violation.

\begin{table}[tbp]
\centering
\caption{Artifact-to-claim traceability. Paths are relative to the
repository root of the tagged release.}
\label{tab:conformance}
\small
\begin{tabularx}{\linewidth}{@{}lX@{}}
\toprule
Claim & Where to verify \\
\midrule
Canonical schema & \texttt{schema/genetic\_evidence.shacl.ttl} (three enforced activation shapes: Variant Ascertainment, Mode of Inheritance, and Organism) \\
YAML-to-RDF transform & \texttt{src/python/main/forome/gem/extraction/yaml\_to\_rdf.py} (normalizes the v0/v1 field renames \texttt{key\_phrase}/\texttt{phrase} and \texttt{resolution}/\texttt{target\_resolution}) \\
Corpus validation & \texttt{gem-validate} over \texttt{annotations/}; CI workflow \texttt{.github/workflows/validate.yml} \\
Coverage counts & \texttt{gem-coverage} regenerates \texttt{annotations/coverage.md} \\
Activation-shape tests & \texttt{src/python/test/forome/gem/validation/} \\
UMLS crosswalk & \texttt{data/umls/} (harness \texttt{gem-umls-crosswalk}; decision log \texttt{data/umls/DECISIONS.md}; SN8 regenerated by \texttt{gem-umls-render}) \\
Archived release & Zenodo deposit of the tagged release; metadata in \texttt{.zenodo.json} \\
\bottomrule
\end{tabularx}
\end{table}

\section*{Acknowledgements}

The idea of a semantic model for the analysis of genetic literature
grew out of 2018 conversations with Prof.\ Shamil Sunyaev; without
his early support this work could not have been completed. The
authors are also grateful to colleagues at the Brigham Genomics
Medicine (BGM) program, where much of the thinking about
clinical-genomics evidence curation took shape, and to
Vladimir Seplyarskiy for his meticulous curation of the four
manually annotated papers in the pilot corpus. This work received
no external funding.

\section{Conclusion}
\label{sec:conclusion}

We have introduced a semantic model for genetic evidence from the
biomedical literature: a compact hierarchy of core classes, a
dimensional vocabulary with conditional-activation rules, three of
which are presently machine-enforced in SHACL, and a SHACL encoding
for automated conformance validation. The model is grounded
in an open corpus of six annotated papers and accompanied by a
collaborative workflow that keeps curator-authored reference annotations
distinct from AI-drafted annotations and promotes candidate schema
extensions only when a second independent paper confirms them. The
model is also released with executable annotation and review skills
that let an AI assistant apply it under curator supervision.
Taken together, these elements are offered as a small but concrete
contribution toward trustworthy, AI-ready infrastructure for variant
interpretation: a reference data model and validation schema for
representing genetic evidence from the biomedical literature.

    \bibliographystyle{plainnat}
    \bibliography{references}

\end{document}


\maketitle
    \setcounter{tocdepth}{2}   %
    \tableofcontents

    \clearpage
    \listoftables

    \clearpage
    \part*{Supplementary Notes}
    \addcontentsline{toc}{part}{Supplementary Notes}
    
\section{Background and Related Work}
\label{sec:supp-background}

Scientific and clinical evidence has been the subject of sustained standards
work in clinical informatics, biomedical ontology, and genomic knowledge
representation. Our schema lies at the intersection of these communities. It
is oriented toward downstream decision-making, grounded in formal constraints
and explicit provenance, and treats target resolution, including variant,
gene, region, and whole-genome aggregate, as a first-class concern while
relying on GA4GH standards for sequence identifiers.

This note positions the schema relative to seven families of standards and
identifies the coverage gap that motivated the work. A first term-level
alignment is provided in \Cref{sec:supp-crosswalk}. Completing the coverage
and formalizing the mappings in OWL remain future work.

\paragraph{FHIR Evidence Resources.}
The HL7 FHIR Evidence, EvidenceVariable, and EvidenceReport
resources~\cite{fhir_evidence} provide structured representations of
clinical-trial-style evidence, with explicit fields for population,
intervention, comparator, outcome, and statistical measures. They were
designed with systematic-review and guideline-development workflows in mind,
where evidence concerns an intervention tested in a population. This framing
does not fit basic-science and pre-clinical genetic evidence cleanly. A mouse
knockout study, for example, does not have an ``intervention'' in the
clinical-trial sense, and a coding-variant case--control study has a target,
the variant, that is not the same kind of object as a drug or procedure. Our
schema shares FHIR Evidence's commitment to structured, dimensional
description but replaces its clinical-trial frame with a vocabulary suited to
genetic claims.

\paragraph{Evidence Codes: ECO and SEPIO.}
The Evidence and Conclusion Ontology (ECO)~\cite{eco2019} provides a large
hierarchical vocabulary of evidence types, including inference methods, assay
types, and source-document categories. It is organized within the Open
Biological and Biomedical Ontologies (OBO) framework and is widely used in
Gene Ontology annotation. The Scientific Evidence and Provenance Information
Ontology (SEPIO)~\cite{sepio} complements ECO with machine-readable
representations of statements, qualifiers, and provenance relationships. Our
Method and Measurement Target dimensions overlap substantially with areas
already covered by ECO and SEPIO. Our long-term goal is to ground those
dimensions in ECO and SEPIO identifiers rather than maintain a separate
vocabulary.

\paragraph{GA4GH Genomic Knowledge Standards.}
The most relevant comparison within the GA4GH Genomic Knowledge Standards
effort is the Variant Annotation (VA) Specification~\cite{ga4gh_va}. VA is a
broad information model for structured statements about variants, organized
around a \texttt{Statement} class whose instances carry a proposition,
direction of support, strength, classification, evidence lines, and
contribution metadata. VA is information-centric: its top-level abstraction,
\texttt{InformationEntity}, represents what is known about a variant rather
than the process by which primary-literature claims become evidence. Our
schema is narrower and complementary. It focuses on evidence drawn from the
primary literature, with the \emph{Relevance and Specificity} and
\emph{Confidence} challenges identified in the main-text introduction as
concrete targets. The Variation Representation Specification
(VRS)~\cite{ga4gh_vrs}, a separate product of the same effort, addresses
canonical variant identity across coordinate systems and is orthogonal to the
present work.

\paragraph{ACMG/AMP variant interpretation.}
The American College of Medical Genetics and Genomics and Association for
Molecular Pathology (ACMG/AMP) guidelines~\cite{acmg2015} specify a
rule-based framework for variant classification from weighted evidence codes
and criteria. Our schema is intended to supply inputs to such frameworks,
rather than replace them: its per-publication evidence items are the kind of
input an ACMG/AMP-compatible classifier would consume. A term-level alignment
of the Method and Credibility dimensions with ACMG/AMP evidence codes remains
future work.

\paragraph{Upper ontologies: IAO and OBI.}
The Information Artefact Ontology (IAO)~\cite{iao2015} and the Ontology for
Biomedical Investigations (OBI)~\cite{obi2016} provide upper-level
scaffolding for many OBO-compliant biomedical ontologies. IAO supplies
classes such as \emph{information-content-entity} and
\emph{textual-entity}; OBI supplies assay and protocol hierarchies relevant
to the Method dimension. Formal alignment of the model's top-level classes
with IAO and OBI, together with an OWL/LinkML rendering of the schema,
remains future work.

\paragraph{Provenance: PROV-O.}
The W3C PROV Ontology (PROV-O)~\cite{prov_o} provides standard provenance
relationships. Aligning the model with PROV-O remains future work.

\paragraph{Related annotation efforts.}
The Clinical Genome Resource (ClinGen) evidence-curation
framework~\cite{clingen2018} provides a structured workflow for gene--disease
validity and variant-pathogenicity assessments with explicit evidence
rubrics. Our schema shares its commitment to source-anchored claims and
explicit categorization, but operates at a lower level by representing the
individual claims that such frameworks aggregate. The Monarch Disease Ontology
(MONDO)~\cite{mondo} and the Human Phenotype Ontology (HPO)~\cite{hpo}
provide established vocabularies for disease and phenotype terms. We expect
curator-authored HPO terms to populate the schema's phenotype-mapping fields,
as illustrated by the HPO-term suggestions in the Davis, Nelson, and Gupta
case studies (Section~5 of the main paper).

\subsection{Requirements and coverage matrix}
\label{sec:supp-coverage-matrix}

The per-family paragraphs above describe each standard on its own terms.
\Cref{tab:coverage-matrix} takes the complementary perspective: it begins
with ten evidence patterns that a genetic publication may report and asks
whether each pattern can be represented by each standard and by GEM.
The six-paper corpus (\Cref{tab:corpus}) exercises GEM's representation of
the first eight patterns, either natively or through a forced fit. The final
two are not represented in the current model. Cross-publication synthesis is
outside GEM's scope by design, whereas therapeutic-response evidence remains
an open scope question under CE-DU3 (row 9). The corpus exposed therapeutic
response directly through the cited anti-p40 antibody trial in Duerr
(GE-new-1). It also exposed a related within-paper synthesis problem: a
higher-order claim over several items in one paper. Both issues are recorded
under candidate extension CE-DU3.

The matrix is a coverage inventory, not a ranking. The standards were
developed for different purposes, and a pattern marked as not represented is
a limitation only if that pattern lies within the standard's intended scope.

The five standards columns represent the resources in this comparison that
carry evidence content. ECO and SEPIO are shown separately, whereas the
ClinGen and ACMG/AMP frameworks share one column. IAO, OBI, and PROV-O are
omitted because they are orthogonal to all ten patterns. They provide
upper-level classes, assay hierarchies, and provenance relations that the
represented models may reuse, but do not themselves represent these patterns.

Cell values are interpreted relative to each standard's role. For an
information model, including FHIR Evidence, GA4GH VA, and SEPIO,
\cfull{} indicates that a class or published profile represents the
pattern. \chalf{} indicates that the pattern is representable only through
the standard's extension mechanism, such as a custom VA profile, a FHIR
extension, or an external predicate vocabulary; only by reference to a
resource outside the module named in the column; or only in one aspect.
\cnone{} indicates that the model has no representation for the pattern.
For ECO, \cfull{} indicates that a term for the pattern exists, \chalf{}
that only a generic term applies, and \cnone{} that no suitable term was
identified.

The ClinGen gene--disease validity framework~\cite{strande2017} and the
ACMG/AMP variant-interpretation guidelines~\cite{acmg2015} share one
column because they score evidence at complementary levels: ClinGen at the
gene level (gene--disease validity) and ACMG/AMP at the variant level
(variant pathogenicity). The column is read as their union. \cfull{}
indicates that the pattern is a scored evidence category in the framework
that operates at the pattern's level, or in both frameworks when it is
scored at both levels (rows 1, 4, and 10); in row 10 the scored category
is the classification the criteria combine into. The row note names the
scoring framework or frameworks. The absence of a criterion in one
framework is not counted as a gap when the pattern lies outside its level,
as a gene-level claim does for ACMG/AMP (rows 2, 5, and 6). \chalf{}
indicates partial or indirect treatment, including a required input that
is not itself scored (row 7) or treatment by a ClinGen product other than
the validity framework (rows 3, 8, and 9). \cnone{} indicates that neither
framework provides a corresponding scored category. For GEM, \cfull{} indicates that
the pattern is represented natively and was annotated in the corpus without
a forced fit. \cforce{} indicates that the pattern was annotated only
through a forced fit recorded as a candidate extension, identified by a CE
code in \fpath{schema/EXTENSIONS.md}. \cnone{} indicates that the pattern
is not represented. In rows 9 and 10 the pattern belongs to what a genetic
publication may report, so GEM's open circle marks a scope boundary rather
than a limitation: settled by design in row 10, left open pending CE-DU3 in
row 9. In any column, \cna{} indicates that the pattern
is outside the standard's subject matter altogether, so its absence is not
a coverage gap.
Every identifier cited in the row notes was verified against the relevant
source vocabulary or specification at the time of writing.

\begin{table}[H]
  \centering
  \footnotesize
  \caption[Requirements and coverage matrix]{Ten evidence patterns across
  five standards and GEM. \cfull{} = represented natively; for ECO, a suitable
  term exists; for the combined ClinGen/ACMG column, the framework at the
  pattern's level (gene or variant) provides a scored category or resulting
  classification. \chalf{} =
  represented only through an extension mechanism, only by reference to a
  resource outside the column's module, or only in part; for ECO, only a
  generic term applies; for GEM, a
  forced fit is recorded as a candidate extension. \cnone{} = not
  represented. \cna{} = outside the standard's subject matter. The
  \emph{remaining gap} column is relative to GEM and names a candidate
  extension where one exists. Row notes follow the table. FHIR = HL7 FHIR
  Evidence R5; VA = GA4GH Variant Annotation 1.0 with VRS 2.0 and Cat-VRS
  1.0; ClinGen/ACMG = the ClinGen gene--disease validity framework, with
  other ClinGen products where noted, and the ACMG/AMP
  variant-interpretation guidelines.}
  \label{tab:coverage-matrix}
  \setlength{\tabcolsep}{4pt}
  \begin{tabularx}{\linewidth}{@{}>{\raggedright\arraybackslash}p{0.30\linewidth} c c c c c c >{\raggedright\arraybackslash}X@{}}
    \toprule
    Evidence pattern & FHIR & VA & SEPIO & ECO
      & \begin{tabular}[b]{@{}c@{}}ClinGen/\\ACMG\end{tabular} & GEM
      & Remaining gap (GEM) \\
    \midrule
    1. Mendelian pedigree: variant co-segregates with disease in a family
      & \chalf & \cfull & \chalf & \chalf & \cfull & \cfull
      & no structured segregation strength (meioses, LOD); activation scope for Variant targets (CE-DU4) \\
    2. Case--control burden: gene-level aggregate of rare variants
      & \chalf & \chalf & \chalf & \chalf & \cfull & \cfull
      & cohort composition (CE-IN5) \\
    3. Common-variant association: GWAS discovery, replication, fine mapping
      & \chalf & \chalf & \chalf & \chalf & \chalf & \cfull
      & direction of effect (CE-DU1); activation scope (CE-DU4); no discovery--replication link \\
    4. Functional assay: variant or gene effect measured \emph{in vitro}
      & \chalf & \chalf & \chalf & \cfull & \cfull & \cfull
      & protein-subdomain resolution (CE-J1); assay type recorded as free text \\
    5. Animal model: knockout or conditional-knockout phenotype
      & \chalf & \chalf & \chalf & \cfull & \cfull & \cfull
      & HPO-only phenotype mapping; rescue not distinguished \\
    6. Gene--gene relation: X inhibits, binds, or regulates Y
      & \cnone & \chalf & \chalf & \cfull & \cfull & \cforce
      & relation vocabulary (CE-J3); interaction as target (CE-J2) \\
    7. Variant identity across coordinate systems (protein, cDNA, genomic)
      & \chalf & \cfull & \cna & \cna & \chalf & \cforce
      & cross-level description (CE-N1); transcript coordinates (CE-C1); VRS/HGVS binding \\
    8. Polygenic score: construction and validation of a genome-wide aggregate
      & \chalf & \chalf & \chalf & \chalf & \chalf & \cforce
      & score machinery (CE-IN1 to CE-IN6) \\
    9. Therapeutic response: intervention tested in patients
      & \cfull & \cfull & \chalf & \cfull & \chalf & \cnone
      & no intervention construct; translational scope open (CE-DU3) \\
    10. Cross-publication synthesis: evidence lines combined into a classification
      & \cfull & \cfull & \cfull & \cfull & \cfull & \cnone
      & out of scope by design \\
    \bottomrule
  \end{tabularx}
\end{table}

\paragraph{Row notes.}
Paper keys follow \Cref{tab:corpus}; item identifiers (GE-$n$) follow the
released annotations.

\begin{enumerate}
  \item \emph{Pedigree segregation.} FHIR can represent the family as a
    \texttt{Group} population and the segregation statistic as an ad hoc
    \texttt{statistic}, but it has no dedicated segregation construct. VA
    covers this pattern through the ACMG 2015 Variant Pathogenicity Evidence
    Line profile, in which co-segregation (PP1) is a criterion that an
    evidence line may carry. SEPIO provides an assertion and evidence-line
    frame, but the segregation content must come from a domain model, as in
    the SEPIO-based ClinGen interpretation model. ECO has no term specific
    to co-segregation, pedigree, or linkage evidence; the only applicable
    term is the general ECO:0005613, \emph{inference by association of
    genotype from phenotype}, whose definition covers alleles linked to
    Mendelian disease and whose usage note names segregation of genetic
    markers (the same term credited in rows 2, 3, and 8). ClinGen
    scores segregation as case-level evidence~\cite{strande2017}, and
    ACMG/AMP codes it as PP1~\cite{acmg2015}. GEM represents the pattern
    natively through Mode of Inheritance and Mendelian Segregation, as
    exercised by Gupta GE-1 (X-linked recessive) and Davis GE-5
    (autosomal recessive). Both are Variant-target items, so they populate
    the Mendelian dimensions outside the documented \dimval{Human
    Genetics}~$+$~\dimval{Gene} activation condition; this activation-scope
    gap is logged as CE-DU4 (row 3). Because Mendelian Segregation is
    Boolean, segregation strength (meioses, LOD) has no structured field;
    pedigree size is recorded only as free text (Gupta GE-1).

  \item \emph{Case--control burden.} FHIR represents the study design and
    statistic natively (\texttt{studyDesign}, \texttt{statistic}), but the
    gene-level aggregate exposure must be coded ad hoc. VA 1.0 has
    no gene-level proposition among its base profiles. Its nearest study
    result, Cohort Allele Frequency, is variant-level, so a custom profile
    is required. ECO offers only the generic ECO:0005613. ClinGen scores
    case--control evidence, including aggregate variant
    analysis~\cite{strande2017}. ACMG/AMP classifies single variants and
    has no gene-level burden criterion; its case--control criterion, PS4,
    concerns the prevalence of one variant. The cell therefore reflects
    ClinGen alone. GEM represents the pattern natively in
    Davis GE-1, with Population Genetics and Human Genetics, Statistical
    Genetics, and a Gene target. Case and control counts are recorded only as
    free text (Davis GE-1, \texttt{special\_considerations}), not as
    structured fields; a structured cohort sub-block is candidate extension
    CE-IN5.

  \item \emph{Common-variant association.} FHIR and VA have the same
    limitations described in row 2. Neither distinguishes a genome-wide
    discovery stage from a replication stage, and VA 1.0 has no population-association
    proposition among its base profiles. ECO has no genome-wide-association or replication term
    beyond ECO:0005613. The ClinGen validity framework scores Mendelian
    gene--disease evidence and does not score complex-trait association.
    ACMG/AMP, the framework at the variant level, does not score it either.
    The guidelines state that they are not intended for variants in genes
    associated with multigenic non-Mendelian complex
    disorders~\cite{acmg2015}; their one case--control criterion, PS4, is
    written for rare variants in Mendelian genes (a relative risk or odds
    ratio above 5.0, with a fallback for very rare variants absent from
    controls), an effect size that common-variant associations rarely
    approach; and a common allele itself is treated as benign evidence by
    the population-frequency criteria BA1 and BS1. Within the two
    frameworks the pattern is therefore unscored. ClinGen's Low
    Penetrance/Risk Allele Working Group has, however, issued separate
    recommendations for curating, classifying, and reporting common risk
    alleles~\cite{schmidt2024}, a ClinGen product outside the validity
    framework; by the convention applied to PRS-RS in row 8, this earns the
    half circle.
    GEM represents the pattern natively through the Method hierarchy:
    GWAS, Association Study, and Fine Mapping under Statistical Genetics.
    This is exercised by Duerr GE-1, GE-2, and GE-4. The same paper exposed
    the missing direction-of-effect dimension for a protective allele
    (CE-DU1) and, through the family-based transmission test in GE-3, an
    activation gap for Human Genetics with a Variant target (CE-DU4).

  \item \emph{Functional assay.} FHIR has no population--intervention frame
    for a cell-based assay; the assay system would have to be cast as a
    population. VA covers the variant-level case through the Experimental
    Variant Functional Impact proposition; its companion study-result
    profile is scoped to multiplexed assay scores, and gene-level assays
    such as the Jossin binding and phosphorylation experiments require a
    custom profile. ECO:0000181, \emph{in vitro assay evidence}, names the
    evidence type generically; ECO's assay-specific terms sit in sibling
    branches such as \emph{direct assay evidence} (ECO:0000002) rather
    than beneath it. ClinGen scores experimental evidence, including biochemical
    function and functional alteration~\cite{strande2017}; ACMG/AMP codes
    it as PS3 or BS3. GEM represents this pattern natively, as illustrated
    by Nelson GE-1 to GE-3 and the \emph{in vitro} Jossin items. Jossin
    exposed the missing protein-subdomain resolution (CE-J1). The assay
    itself is described only at the Method level, \emph{In Vitro}. The
    crosswalk maps that value to ECO:0000181 at the same granularity
    (\cref{sec:supp-crosswalk}); grounding assay type in ECO's
    assay-specific terms or in OBI assay classes remains future work.

  \item \emph{Animal model.} FHIR can cast the genotype as an exposure and
    the animal cohort as a \texttt{Group}, but it has no model-organism
    frame. VA has no animal-model proposition; the study can enter only as
    an evidence item behind a custom profile. ECO provides detailed
    evidence-type coverage through ECO:0000179,
    \emph{animal model system study evidence}; ECO:0001091,
    \emph{knockout phenotypic evidence}; ECO:0001030,
    \emph{conditional knockout evidence}; and ECO:0000013,
    \emph{transgenic rescue experiment evidence}. ClinGen scores
    non-human model organisms and rescue as experimental
    evidence~\cite{strande2017}. A knockout phenotype is a gene-level claim
    outside ACMG/AMP's variant-level scope; PS3 and BS3 admit \emph{in
    vivo} functional studies only as evidence of the effect of the variant
    under interpretation on its gene or gene product. The cell therefore
    reflects ClinGen alone. GEM represents the pattern natively with
    the conditional Organism and Knockout Type dimensions, exercised in
    full by Jossin GE-1, a conditional mouse knockout with organismal-scale
    evidence and the corpus's only knockout item. The Davis zebrafish
    (GE-2, a morpholino knockdown with rescue) and rat (GE-4) items
    exercise Organism but not Knockout Type. The
    phenotype-mapping fields currently expect HPO terms, so they do not bind
    model-organism phenotype vocabularies, and rescue is not distinguished
    from other \emph{in vivo} evidence.

  \item \emph{Gene--gene relation.} FHIR Evidence has no frame for a
    relation between two molecular entities. VA includes a \texttt{Gene}
    domain entity but no interaction proposition. A SEPIO assertion can
    carry the relation through a predicate from the Relation Ontology.
    ECO:0000021, \emph{physical interaction evidence}, covers the evidence
    type. ClinGen scores protein interaction as experimental evidence in
    its Function category, alongside the biochemical function evidence of
    row 4~\cite{strande2017}, although the relation serves a gene--disease
    claim rather than standing as a claim of its own. ACMG/AMP has no
    corresponding criterion: a gene--gene relation is a gene-level claim
    outside its variant-level scope, and a variant's effect on an
    interaction would enter only as generic functional evidence under PS3
    or BS3. The cell therefore reflects ClinGen alone. GEM has a Gene
    Relation dimension, of whose
    three current values the corpus exercised one: X inhibits Y in Jossin
    GE-5. The physical-binding relation in GE-4 matched no value and was
    left unpopulated (CE-J3, which also proposes a trafficking-regulation
    value), and the LLGL1--N-cadherin interaction targeted by GE-6 was a
    forced fit as a Gene target (CE-J2, which proposes Interaction and
    Complex values).

  \item \emph{Variant identity across coordinate systems.} This pattern is
    outside the FHIR Evidence module. FHIR handles it through genomics
    reporting resources that an \texttt{Evidence} resource can reference.
    VA delegates variant representation to VRS~\cite{ga4gh_vrs} and
    Cat-VRS: a VRS \texttt{Allele} sits on a genomic, transcript, or protein
    sequence reference and can carry HGVS \texttt{g.}, \texttt{c.}, or
    \texttt{p.} expressions, and the Cat-VRS \texttt{CanonicalAllele} and
    \texttt{ProteinSequenceConsequence} classes group congruent alleles
    across assemblies, transcripts, and proteins, a linkage that VRS itself
    confines to a single sequence context. SEPIO and ECO do not address
    variant identity. Neither curation framework scores identity, but both
    require it as input: ACMG/AMP mandates HGVS descriptions against a
    versioned reference sequence~\cite{acmg2015}, and ClinGen's curation
    interfaces identify a variant by ClinVar or ClinGen Allele Registry
    identifier, whose canonical allele links the genomic, transcript, and
    protein forms. GEM relies on GA4GH identifiers by design, but its
    Resolution values currently presuppose genomic coordinates. The Nelson
    insertion allele, a 12-bp insertion in cDNA coordinates but a
    single-nucleotide substitution in genomic coordinates, was a forced fit
    under Resolution \dimval{Variant} and was recorded as CE-N1, resolved by
    reference to HGVS and VRS rather than by a new value. A
    transcript-coordinate Resolution value (CE-C1) was surfaced by curator
    review during the UMLS crosswalk and has no corpus instance yet.
    Binding target identifiers to VRS or HGVS is planned rather than
    implemented.

  \item \emph{Polygenic score.} FHIR can represent a risk-prediction study
    with the score as a coded variable, but it has no dedicated score
    construct. VA 1.0 and VRS provide no entity for a genome-wide aggregate,
    so a custom profile would be required. ECO offers only the generic
    computational term ECO:0007672 and association term ECO:0005613.
    ClinGen's relevant contribution is the PRS-RS reporting
    checklist~\cite{wand2021}, not a curation rubric. In GEM, all four
    Inouye items are forced fits: Target Type and Resolution both carry the
    placeholder value \dimval{Variant}, and Variant Ascertainment is set to
    the documented sentinel. These forced values motivated CE-IN1 to
    CE-IN3; the same annotation surfaced CE-IN4 to CE-IN6 from its
    epidemiological outcome measures, free-text cohort description, and the
    score's status as a derived artifact. All six await a second
    polygenic-score paper under the two-papers rule.

  \item \emph{Therapeutic response.} This pattern fits FHIR Evidence's
    native population--intervention--comparator--outcome frame. VA covers
    it through the Variant Therapeutic Response proposition. ECO:0000180,
    \emph{clinical study evidence}, covers the evidence type. ClinGen
    addresses therapeutic response through actionability curation rather
    than the gene--disease validity framework. GEM has no intervention
    construct. The cited anti-p40 antibody trial in Duerr, GE-new-1, was
    recorded with Knowledge Domain left unresolved by curator decision rather
    than forced into a genetic domain. It is the second CE-DU3 instance in
    the same paper; the first is the authors' therapeutic-strategy
    recommendation, a higher-order claim over several of the paper's items.
    Two instances in one paper do not satisfy the two-papers rule, so CE-DU3
    remains a candidate whose proposal leaves open whether to add a
    higher-order claim construct or to declare translational claims out of
    scope.

  \item \emph{Cross-publication synthesis.} FHIR \texttt{Evidence} carries
    a synthesized result natively (a rated \texttt{certainty};
    \texttt{synthesisType}, the method used to combine studies; and
    \texttt{numberOfStudies} under \texttt{statistic.sampleSize}), with
    \texttt{EvidenceReport} as the
    container that composes the contributing resources. Combining evidence
    lines into a conclusion is likewise the purpose of VA
    \texttt{Statement}s with evidence lines, including ACMG classification
    profiles; of SEPIO evidence lines; and of the ClinGen and ACMG/AMP
    rubrics. ECO covers the evidence type through ECO:0000212,
    \emph{combinatorial evidence}, and its descendant ECO:0007012,
    \emph{combinatorial experimental and curator inference evidence}; as
    an evidence-type vocabulary it names the combined evidence, not the
    resulting classification. GEM does not synthesize evidence by design.
    The within-paper form of this pattern, a higher-order claim over several
    of a paper's own items, is the open candidate CE-DU3 (row 9). GEM
    represents the per-publication evidence items
    consumed by such frameworks, while the reasoning layer that combines
    them is developed separately~\cite{our_preprint}.
\end{enumerate}

Read by column, the matrix clarifies GEM's contribution. Rows 1 through 5
cover common patterns in basic-science genetic publications. Under the
operational definitions used in this coverage inventory, GEM is the only
resource in this comparison that represents all five patterns natively
within one common, genetics-specific dimensional profile. Other resources
require an extension mechanism because they were designed for a different
primary purpose, provide an evidence-type vocabulary that names the evidence
without structuring the claim, or, in the case of the ClinGen and ACMG/AMP
frameworks, score Mendelian gene- and variant-level evidence as inputs to a
classification while leaving common-variant association to a separate
risk-allele product (row 3).

Rows 6 through 8 identify patterns that the pilot exposed as forced fits and
therefore as documented areas for extension. Row 10 is outside GEM's scope
by design. Row 9, therapeutic response, is not represented in the current
model, but CE-DU3 leaves open whether translational claims should become part
of the model's scope or should be left to standards designed for
intervention evidence.

\paragraph{The coverage gap.}
None of these standards was designed to treat the heterogeneous evidence
patterns of basic-science and pre-clinical genetic publications as a
first-class concern. A Llgl1 conditional-knockout study, for example, may
report molecular, cellular, and clinical phenotypes in the same paper. A
polygenic-score study may aggregate millions of variants into one score. Each
existing standard captures part of what such papers report, but none captures
the whole. The schema introduced in the main paper aims to represent this
heterogeneity in a form that is compact enough for manual annotation,
machine-checkable enough for automated validation, and flexible enough to
accommodate the evidence patterns produced by the literature.

\section{The annotation and review protocol suite}
\label{sec:supp-protocol-suite}

The annotation workflow described in the main paper has been consolidated
into a versioned, openly published protocol suite available in the project
repository.\footnote{\url{https://github.com/ForomePlatform/genetic-evidence-model/tree/master/protocols}}
The suite distinguishes what a well-formed annotation or review must contain
from how it is performed, either manually or with AI assistance. It governs
both pilot tracks: the curator-authored manual track and the AI-drafted
track.

A mode-agnostic annotation protocol (\texttt{PROTOCOL.md}) defines the
requirements for a valid annotation, including classes, dimensions, source
anchoring, and flag semantics. Two mode-specific specializations adapt this
protocol for autonomous drafting (\texttt{PROTOCOL\_AUTONOMOUS.md}) and
interactive, curator-in-the-loop work. A separate review protocol
(\texttt{REVIEW\_PROTOCOL.md}, with an interactive variant) governs
assertion-by-assertion review of drafted annotations. A labeling-examples
document records worked cases for recurring judgment calls. These written
protocols are also packaged as executable skills
(\cref{sec:supp-skills}), one for annotation and one for review, so that the
same rules can be applied by a human curator or by an AI assistant under
curator supervision.

\subsection{Manual-track protocol}

In the \emph{manual track}, four of the six publications, Jossin 2017,
Davis 2011, Nelson 1992, and Gupta 2015, were annotated by Vladimir
Seplyarskiy, a researcher in human and population genetics specializing in
human germline mutation and mutational processes. He is an Assistant
Professor at UT~Southwestern and was formerly affiliated with Harvard Medical
School and Brigham and Women's Hospital. This expertise covers the
human-genetics and population-genetics knowledge domains represented in the
corpus.

For each paper, the curator:

\begin{enumerate}
  \item read the PDF in full and marked relevant passages with highlights
        and short callouts;
  \item exported those annotations with a helper script to a structured JSON
        record containing page, offset, highlighted text, and free-text
        callout fields;
  \item constructed a one-row summary table containing high-level judgments
        about knowledge domain, method, credibility, resolution, mode of
        inheritance, and related dimensions; and
  \item mapped the highlights and summary row to the schema, producing a
        YAML file conforming to the \texttt{GeneticEvidence} model.
\end{enumerate}

When this conversion required normalization, the change was recorded inline
as a \texttt{normalization\_note} with
\texttt{type: ai\_normalization}, while preserving the original text. For
example, the free-text value ``localisation and protein expression'' was
mapped to the enumeration values \texttt{LOCALIZATION} and
\texttt{EXISTENCE}. These notes allow later audits to compare the original
and normalized values and support subsequent ontology alignment.

All manual-track annotations were validated against the SHACL schema
(\fpath{schema/genetic_evidence.shacl.ttl}). Violations, such as a missing
always-required dimension or a conditional dimension whose activation
condition held but whose value was absent, had to be resolved before an
annotation was admitted to the corpus.

The four manual annotations are treated as the curator-authored
\emph{reference annotations} for this work. They are not a gold standard in the statistical sense, because the
corpus has only one annotator. They are, however, the authoritative reference
against which the AI-drafted track is evaluated.

\subsection{AI-drafted-track protocol}

In the \emph{AI-drafted track}, an AI assistant annotated two publications,
Duerr 2006 and Inouye 2018, directly from their PDFs under the same schema.
The four manual annotations served as worked examples.

For each paper, the assistant received:

\begin{itemize}
  \item the schema (\fpath{schema/dimensions.md} and the SHACL file);
  \item the four manual annotations;
  \item the source PDF; and
  \item the autonomous-annotation skill (\cref{sec:supp-skills}).
\end{itemize}

The assistant produced draft YAML annotations in the same format as the
manual annotations. When it judged a mapping uncertain, or concluded that a
judgment required human review, it recorded the issue inline using one of two
flag types:

\begin{itemize}
  \item \texttt{ai\_uncertainty}, for a mapping judged unreliable and
        referred for a human decision; and
  \item \texttt{ai\_assumption}, for a judgment made without explicit
        textual support and requiring reconsideration at review rather than
        silent acceptance.
\end{itemize}

AI-drafted annotations are not accepted without review. One of the authors,
a curator with expertise in genomic data curation and the evidence model
rather than in human or population genetics, reviewed each annotation
assertion by assertion under the review skill. The review assessed:

\begin{itemize}
  \item faithful representation of the source text;
  \item conformance to the schema, including conditional-activation rules;
        and
  \item the appropriateness of any \texttt{ai\_uncertainty} or
        \texttt{ai\_assumption} flags.
\end{itemize}

Corrections were applied directly to the YAML, and every change and ruling
was logged under \fpath{annotations/reviews/}. \Cref{tab:ai-review} summarizes the
resulting verdicts for each \texttt{GeneticEvidence} item, and
\cref{sec:supp-ai-review} analyzes them qualitatively. In all cases, the \emph{curator}, not
the model, is the authority for the recorded annotations.

\subsection{Source anchoring and validation}

A central epistemic commitment of the workflow is that every
\emph{assertion} must be independently auditable against its source.

\begin{itemize}
  \item Each assertion in a \texttt{GeneticEvidence} item carries a
        \texttt{source\_span} containing a page number and a concise key
        phrase, typically fewer than fifteen words, sufficient to locate the
        passage by search in the PDF.
  \item The phrase-length limit is intended to respect fair-use norms for
        copyrighted publisher material while keeping verification practical.
  \item The SHACL schema treats \texttt{source\_span} as mandatory. An
        annotation with any assertion lacking a source anchor is rejected at
        validation time.
\end{itemize}

Across the six-paper pilot, all 95 assertions carry source spans. This is a
property of the validation requirement, not an observed outcome. When a
verbatim key phrase could not be reconstructed from a public PDF, for example
because a value appeared in a publisher PDF table not included in the
repository, the annotation records this explicitly with a
\texttt{source\_span\_is\_paraphrase} flag and an explanatory note. Future
curators can then refine the anchor against the original file.

\subsection{Provenance and protocol versions}

Two provenance points are important for interpreting the corpus.

First, the pilot corpus described in the main paper was initially annotated
under an earlier procedure, ``protocol~v0.'' The suite described here is its
subsequent, generalized formalization. The pilot annotations remain valid
under the current protocol. Some artifacts, such as slightly different
span-selection conventions in the earliest AI-drafted annotations, are
recorded as forward-looking notes rather than retrofitted edits.
Two such artifacts are field-name changes. Protocol~v0 used
\texttt{key\_phrase} for the source-span phrase and \texttt{resolution} for
the target resolution, whereas v1 uses \texttt{phrase} and
\texttt{target\_resolution}. The corpus retains both spellings. Before RDF
validation, the YAML-to-RDF transformation
(\fpath{src/python/main/forome/gem/extraction/yaml_to_rdf.py}) normalizes
them to the RDF properties \texttt{gem:phrase} and \texttt{gem:resolution},
so \texttt{gem-validate} treats the two spellings identically; the legacy
names are retained solely for backward compatibility.

Second, the AI-drafted annotations were produced with Claude Opus~4.7 for
Duerr and Claude Opus~4.8 for Inouye, and the assertion-by-assertion review
was conducted by the human curator interacting with Claude Opus~4.8 under
the executable review skill. These implementation details are reported for
reproducibility and are not central to the scientific contribution.
AI-generated drafts and AI-assisted review suggestions were subject to
final adjudication by the human curator; the reviewed annotations are
human-approved records, and no AI system was treated as the authority for
any scientific or curation decision.

\section{Executable skills}
\label{sec:supp-skills}

Because the term may be unfamiliar outside agent-based AI tooling, we briefly
define an executable \emph{skill}. A skill is a version-controlled package of
natural-language instructions and supporting resources, such as reference
files, templates, and worked examples. An AI assistant loads the package on
demand to perform a defined task in a repeatable way~\cite{anthropic_skills}.

A skill is the executable counterpart of a written protocol. A protocol tells
a human reader what a task requires; a skill expresses those requirements in a
form that an AI assistant can apply directly. In practice, a skill is a
folder containing a top-level instruction file, conventionally
\texttt{SKILL.md}, and the resources referenced by that file. The assistant
reviews the skills available to it, loads the one relevant to the task, and
follows its instructions. Its behavior is therefore governed by the
versioned package rather than improvised anew for each session.

The general mechanism is not specific to a particular model or vendor, and a
cross-vendor specification is emerging.\footnote{\url{https://agentskills.io}}
The two skills accompanying this work are implemented as Claude Agent Skills.

The annotation skill packages the annotation protocol, and the review skill
packages the review protocol (\cref{sec:supp-protocol-suite}). They allow an
AI assistant to apply the same rules that guide a human curator, under curator
supervision. Both skills are published with the protocols in the project
repository.\footnote{\url{https://github.com/ForomePlatform/genetic-evidence-model/tree/master/skills}}
Their purpose in this work is reproducibility and reuse, not autonomy. The
curator remains the authority for every recorded annotation, as described in
\cref{sec:supp-protocol-suite}.

\section{AI-track review outcomes}
\label{sec:supp-ai-review}

This note reports the review outcomes for the two AI-drafted annotations:
Duerr et al.\ 2006, a conventional genome-wide association example, and
Inouye et al.\ 2018, a polygenic-score case that exposes several model
limitations.
Each draft was reviewed assertion by assertion against
\texttt{REVIEW\_PROTOCOL}~v1.0 by the human curator, who interacted with
an AI assistant operating under the review skill.
AI-generated drafts and
AI-assisted review suggestions were subject to final adjudication by the
human curator; the reviewed annotations are human-approved records, and no
AI system was treated as the authority for any scientific or curation
decision.

The complete review logs, including every flag and ruling, are available in
the repository under \fpath{annotations/reviews/}.
\Cref{tab:ai-review} summarizes the verdicts for the drafted items.

\begin{table}[H]  %
  \centering
  \small
  \begin{tabular}{lcccccc}
    \toprule
    Paper & Drafted & Approved & With flags & Edited & Rejected & Added on review \\
    \midrule
    Duerr 2006   & 6 & 0 & 2 & 4 & 0 & 1 (approved) \\
    Inouye 2018  & 3 & 1 & 1 & 1 & 0 & 1 (edited) \\
    \midrule
    Total        & 9 & 1 & 3 & 5 & 0 & 2 \\
    \bottomrule
  \end{tabular}
  \caption[Per-paper AI-track review verdicts]{Per-paper review verdicts
  for AI-drafted items. ``With flags'' denotes approval with non-blocking
  flags and no edits. ``Added on review'' denotes evidence omitted from the
  draft and entered by the reviewer as a new item.}
  \label{tab:ai-review}
\end{table}

The Duerr results reflect the \texttt{v0} annotation. The Inouye results,
and all related analyses in this note, use the \texttt{v1} re-annotation
produced under the protocol refined during the Duerr review. The original
five-item \texttt{v0} Inouye draft is retained in the repository for
comparison.

The drafts were produced under different protocol versions. The later Inouye
draft required fewer edits: two of its three items were accepted without
edits and one was edited, whereas four of Duerr's six items were edited.
This comparison is only suggestive, not controlled, because the papers differ
in both genre and difficulty. It is nevertheless consistent with the
provenance record: all thirty-seven cited Inouye spans were verbatim and
within the length limit, whereas the earlier Duerr draft contained notation
drift and one incorrect page citation. These issues are retained as
forward-looking notes because they predate the current rules.

A second pattern concerns omitted evidence. The reviewer-added items were not
granular association findings, which the drafts captured well. Instead, they
were higher-level or interpretive items: in Inouye, the paper's headline
claim that combining component scores outperforms each component; and in
Duerr, a cited anti-p40 antibody trial supporting the therapeutic-target
argument. The drafts were therefore most reliable for specific, directly
stated claims and less reliable for synthetic conclusions and cited
translational evidence that required cross-item interpretation.

Edits to existing items primarily concerned schema-level judgment rather than
the factual content of assertions. They included Knowledge Domain assignment,
distinguishing family-based from allele-frequency evidence; Method placement,
distinguishing targeted replication from a new genome-wide scan; Credibility
calibration, lowering the score-combination item from high to medium because
the paper gave no stated mechanism; and the applicability of Variant
Ascertainment to a polygenic-score target. In the last case, the reviewer
marked the dimension as not applicable and recorded the resulting constraint
as support for a candidate extension. No item was rejected, and no item was
found to misrepresent its source.

The revision also affects the promotion record. Because Variant Ascertainment
is no longer exercised by Inouye, its promotion rests on its initial proposal
in Davis and its independent reuse in Nelson and Duerr, which still satisfies
the two-papers rule. Updating the promotion record is a separate curation
step and is not undertaken here. Both reviews recommend a second pass over
the changed items because each paper gained a new item and several existing
items were edited.

    \section{Standards crosswalk}
\label{sec:supp-crosswalk}

This section extends the class-level FHIR alignment in the main paper with
term-level mappings for GEM classes, properties, dimensions, and selected
enumerated values. The mappings cover the Sequence Ontology (SO), the Human
Phenotype Ontology (HPO), NCBITaxon, the Evidence and Conclusion Ontology
(ECO), SEPIO, and the GA4GH Variant Annotation (VA) model. Each mapping was
verified against its source vocabulary.

This standards crosswalk is distinct from the UMLS crosswalk in
\Cref{sec:s-umls-crosswalk}. It aligns GEM constructs with source
vocabularies, whereas that note records concept-level mappings and curation
decisions within the UMLS Metathesaurus.

The \emph{relation} column describes the relation between a GEM construct
and an external term. \emph{Exact} indicates the same scope;
\emph{close} a near match with a material scope difference;
\emph{narrower} that GEM is more specific; \emph{broader} that GEM is more
general; \emph{related} that the external term captures a constituent or
associated aspect of the GEM construct; and \emph{none} that no equivalent
has been identified. For structural correspondences that are not
term-level equivalences, the tables instead use structural labels
(\emph{structural sibling}, \emph{structural parallel}).

The crosswalk prioritizes semantic fidelity over a fixed preference for one
vocabulary. When the most faithful concept is available only in a specialized
vocabulary, the crosswalk retains that exact match rather than substituting a
broader term from a preferred vocabulary. For example,
\dimval{ANIMAL\_GENETICS} maps exactly to NCI Thesaurus
\texttt{C1510895} (\emph{Animal Genetics}), for which MeSH has no
corresponding descriptor. The \emph{relation} field records the nature of
the mapping; it does not express a preference among vocabularies.

A second source-level issue is sense merging within a single UMLS concept.
For \dimval{VARIANT}, UMLS concept \texttt{C0042333}
(\emph{Genetic Variation}) combines two senses. The MeSH descriptor defines
a population-level phenomenon (``genotypic differences observed among
individuals in a population'') and determines the concept's semantic type,
\emph{Natural Phenomenon or Process}. The NCI Thesaurus instead defines an
entity, a deviation in the nucleotide sequence of an individual, and places
the ACMG classification variants (\emph{pathogenic}, \emph{likely
pathogenic}, and \emph{benign}) directly under that concept. Its synonym
list also includes \emph{Sequence Variant}, the exact SO term shown in
\cref{tab:xwalk-dimensions}.

The UMLS mapping is therefore recorded as \emph{close}, rather than
\emph{exact}, because the concept includes the population-level MeSH sense.
We nevertheless accept its semantic type even though it lies outside the
expected axis. The entity-level NCI sense is the relevant genomics sense,
and UMLS provides no generic variant concept that is typed as an entity:
the classified child concepts are typed \emph{Nucleotide Sequence}, but
their generic parent is not.

\begin{table}[H]
  \centering
  \small
  \caption[Crosswalk: structural classes]{Structural correspondences between
  GEM and FHIR R5. The table extends the class-level alignment in the main
  paper with term-level detail.}
  \label{tab:xwalk-structural}
  \begin{tabularx}{\linewidth}{@{}>{\raggedright\arraybackslash}X >{\raggedright\arraybackslash}X l >{\raggedright\arraybackslash}X@{}}
    \toprule
    GEM construct & External term & Vocab. & Relation \\
    \midrule
        \ScientificEvidence{} (evidence item) & \texttt{Evidence} & FHIR R5 & structural sibling; GEM replaces the PICO frame with a genetics-specific dimensional frame \\
        \EvidenceVariable{} (definition and measured value) & \texttt{EvidenceVariable} + \texttt{Evidence.statistic} & FHIR R5 & broader; GEM combines the variable definition and measured value in one class \\
        Credibility (overall rating slot) & \texttt{Evidence.certainty} & FHIR R5 & structural parallel; both provide an overall rating with separable subcomponents \\
    \bottomrule
  \end{tabularx}
\end{table}

\begin{table}[H]
  \centering
  \small
  \caption[Crosswalk: assertion property and class]{Crosswalk for the
  \texttt{assertion} property and the \EvidenceAssertion{} class. The
  property records an object-level, source-anchored claim; the class is a
  meta-predicate over a decision rule.}
  \label{tab:xwalk-assertion}
  \begin{tabularx}{\linewidth}{@{}>{\raggedright\arraybackslash}X >{\raggedright\arraybackslash}X l >{\raggedright\arraybackslash}X@{}}
    \toprule
    GEM construct & External term & Vocab. & Relation \\
    \midrule
    \texttt{assertion} property (source-anchored claim) & \texttt{Statement} & GA4GH VA & narrower; a close, genetics-specific match \\
        \texttt{assertion} property & \texttt{Assertion} (\texttt{SEPIO:0000001}) & SEPIO & narrower \\
        \texttt{assertion} property (false friend) & \texttt{Evidence.assertion} (human-readable summary) & FHIR R5 & none; do not map \\
        \EvidenceAssertion{} class (meta-predicate over a decision rule) & no equivalent & --- & none \\
    \bottomrule
  \end{tabularx}
\end{table}

The \texttt{assertion} property and the \EvidenceAssertion{} class are
distinct despite their similar names. The property records the object-level,
source-anchored claim made by a paper. Its closest structural analogues are
the GA4GH VA \texttt{Statement} and the SEPIO \texttt{Assertion}, not
FHIR's \texttt{Evidence.assertion}, which is a human-readable summary.

The \EvidenceAssertion{} class is a meta-predicate over a decision rule and
is the model's point of contact with the reasoning layer
of~\cite{our_preprint}. We found no close analogue for this decision-rule
meta-predicate in FHIR, GA4GH VA, ECO, or SEPIO. Its lack of an external
analogue is therefore expected rather than a coverage failure. A future
schema revision will rename the property to remove the naming collision; the
pilot annotations will remain unchanged.

\begin{table}[H]
  \centering
  \footnotesize
  \caption[Crosswalk: dimensions and values]{GEM dimensions and selected
  values mapped to external vocabularies. Identifiers were verified against
  their source vocabularies. SO = Sequence Ontology; VA = GA4GH Variant
  Annotation.}
  \label{tab:xwalk-dimensions}
  \begin{tabularx}{\linewidth}{@{}>{\raggedright\arraybackslash}X >{\raggedright\arraybackslash}X l >{\raggedright\arraybackslash}X@{}}
    \toprule
    GEM dimension / value & External term + ID & Vocab. & Relation \\
    \midrule
        Target Type: gene & gene \texttt{SO:0000704} & SO & exact \\
        Target Type: related gene & gene \texttt{SO:0000704} (relation recorded in Gene Relation) & SO & exact \\
        Target Type: variant & sequence\_variant \texttt{SO:0001060} & SO & exact \\
        Target Type: transcript & transcript \texttt{SO:0000673} & SO & exact \\
        Target Type: interval / segment & region \texttt{SO:0000001} & SO & narrower \\
        Mode of Inheritance (axis) & Mode of inheritance \texttt{HP:0000005} & HPO & exact \\
        \quad autosomal dominant & \texttt{HP:0000006} & HPO & exact \\
        \quad autosomal recessive & \texttt{HP:0000007} & HPO & exact \\
        Organism: human & \textit{Homo sapiens} \texttt{NCBITaxon:9606} & NCBITaxon & exact \\
        Organism: mouse & \textit{Mus musculus} \texttt{NCBITaxon:10090} & NCBITaxon & exact \\
        Organism: rat & \textit{Rattus norvegicus} \texttt{NCBITaxon:10116} & NCBITaxon & exact \\
        Organism: zebrafish & \textit{Danio rerio} \texttt{NCBITaxon:7955} & NCBITaxon & exact \\
        Credibility (rating and facets) & \texttt{confidence level} (\texttt{SEPIO:0000187}), via \texttt{has\_confidence\_level} (\texttt{SEPIO:0000167}), on a rated \texttt{Assertion} (\texttt{SEPIO:0000001}) & SEPIO & exact (rating) / narrower (facets) \\
        Method: in vivo (animal model) & animal model system study evidence \texttt{ECO:0000179} & ECO & exact \\
        Method: in vitro & in vitro assay evidence \texttt{ECO:0000181} & ECO & exact \\
        Method: clinical & clinical study evidence \texttt{ECO:0000180} & ECO & exact \\
        Method: statistical genetics & no clean ECO leaf; broadest parent is \texttt{computational evidence} (\texttt{ECO:0007672}); OBI is a candidate vocabulary for study designs (unconfirmed) & ECO / OBI & none (ECO) / broader (OBI) \\
    \bottomrule
  \end{tabularx}
\end{table}

The following table lists dimensions for which the selective review reported
here identified no single external equivalent. These entries are documented
coverage gaps in the present crosswalk, not claims that no relevant term
exists in any vocabulary.

\begin{table}[H]
  \centering
  \small
  \caption[Crosswalk: coverage gaps]{Coverage gaps in the present crosswalk.
  The relation is \emph{none} in every row.}
  \label{tab:xwalk-gaps}
  \begin{tabularx}{\linewidth}{@{}l >{\raggedright\arraybackslash}X@{}}
    \toprule
    GEM dimension & Why no external term \\
    \midrule
    Knowledge Domain & no single external vocabulary represents this axis \\
        Gene Relation (conditional) & GEM-specific gene--gene relations, such as \texttt{X\_regulates\_Y} and \texttt{X\_has\_same\_function\_as\_Y}; RO may cover some relations (unconfirmed) \\
        Phenotype Scale (ordinal) & no external vocabulary provides this ordinal axis; GO biological\_process is a possible future alignment for the molecular and cellular end of the scale \\
        Variant Ascertainment & no single external term captures this dimension \\
        Resolution & represented implicitly by the selected SO target-type term, rather than as an independent external axis \\
        \EvidenceAssertion{} (meta-predicate) & belongs to a reasoning layer beyond the scope of ECO, SEPIO, VA, and FHIR \\
    \bottomrule
  \end{tabularx}
\end{table}

Taken together, the mappings and gaps show two broad patterns. First, the
dimensions that describe what the evidence concerns and which organism it
concerns map directly to established ontologies. The three experimental
Method values also map cleanly. Statistical genetics is the main Method gap:
ECO represents evidence types rather than study designs such as genome-wide
association studies and linkage analyses, and OBI is a candidate vocabulary
for that level of description.

Second, this crosswalk identifies no single external equivalent for Knowledge
Domain, Phenotype Scale, Variant Ascertainment, Resolution, or the Gene
Relation enumeration. These are GEM-specific epistemological axes, rather
than failed attempts to duplicate existing vocabularies. Their partial or
absent coverage in the reviewed standards constitutes a coverage gap in the
present crosswalk and helps motivate a dedicated reference data model.

    \section{Case reports: Duerr 2006 and Inouye 2018}
\label{sec:supp-crep}

The two AI-drafted annotations illustrate contrasting parts of the model's
operating range. The main paper (Section~4) summarizes them; this
note provides the item-level detail omitted there for space.
Duerr is the more conventional case: six of its seven \GeneticEvidence{}
items fit the current schema without strain, and the remaining item raises
an open question about translational scope. Inouye tests the limits of the variant-level model: all four items
require forced-fit assignments, which motivate six candidate extensions.
Assertion-by-assertion review verdicts are given in
\cref{sec:supp-ai-review}; the term-level standards crosswalk is in
\cref{sec:supp-crosswalk}; and the complete annotations are available as
the repository files \fpath{annotations/v0/duerr2006.yaml} and
\fpath{annotations/v1/inouye2018.yaml}.

\subsection{Case Report CR-DU: Duerr et al.\ 2006 (genome-wide association)}
\label{sec:supp-ex-duerr}

Duerr and colleagues identified \textit{IL23R} as an inflammatory bowel
disease gene through a genome-wide association study of non-Jewish patients
with ileal Crohn's disease, followed by replication in a Jewish cohort and
family-based transmission testing in 883 nuclear families. Our annotation
decomposes the paper into seven \GeneticEvidence{} items
(\cref{tab:cr-duerr}).

\begin{table}[H]
  \centering
  \footnotesize
  \caption[Duerr 2006: evidence-item decomposition]{Duerr 2006: the seven
  \GeneticEvidence{} items, with target type, leaf-level Method, and
  Credibility.}
  \label{tab:cr-duerr}
  \begin{tabularx}{\linewidth}{@{}l >{\raggedright\arraybackslash}X >{\raggedright\arraybackslash}p{1.35cm} >{\raggedright\arraybackslash}p{2.5cm} l@{}}
    \toprule
    ID & Claim & Target & Method & Cred. \\
    \midrule
    GE-1 & rs11209026 protective in non-Jewish ileal CD (discovery) & \dimval{Variant} & \dimval{GWAS} & High \\
    GE-2 & rs11209026 protective in Jewish ileal CD (replication) & \dimval{Variant} & \dimval{Association Study} & High \\
    GE-3 & Gln undertransmitted to affected offspring (FBAT) & \dimval{Variant} & \dimval{Transmission Disequilibrium Test} & High \\
    GE-4 & Multiple independent IL23R signals (conditional) & \dimval{Variant} & \dimval{Fine Mapping} & High \\
    GE-5 & No association at IL12RB1/IL23A/IL12B & \dimval{Gene} & \dimval{GWAS} & High \\
    GE-6 & Cited mouse-model and functional support & \dimval{Gene} & \dimval{In Vivo}, \dimval{In Vitro} & High \\
    GE-new-1 & Cited anti-p40 (IL12B) antibody trial & \dimval{Gene} & \dimval{Clinical Evidence} & High \\
    \bottomrule
  \end{tabularx}
\end{table}

Six of the seven items fit the current schema without strain. The seventh,
GE-new-1, is a cited anti-p40 antibody trial added during the
missing-evidence check. It concerns drug-response evidence for which the
model has no fitting genetic Knowledge Domain, revealing a
translational-scope gap. The discovery item, GE-1, is shown in abridged form
below. It illustrates how dimensions, a credibility comment, and
source-anchored assertions are recorded.

\begin{lstlisting}[language=yaml]
id: GE-1
label: "rs11209026 (Arg381Gln) in IL23R confers protection
         against ileal CD in non-Jewish discovery cohort"
knowledge_domain: [POPULATION_GENETICS]
method: [GWAS]
target_type: VARIANT
target: "rs11209026 (c.1142G>A, p.Arg381Gln) in IL23R"
resolution: VARIANT
variant_ascertainment: [OBSERVED_IN_CASES, OBSERVED_IN_CONTROLS]
phenotype_scale: CLINICAL
credibility: HIGH
credibility_comment: >
  Genome-wide significance after Bonferroni correction
  (corrected P = 1.56e-3); large effect (OR = 0.26);
  independent replication follows in GE-2.
candidate_extension_exercised: CE-DU1   # direction of effect
assertions:
  - id: GE-1.A1
    statement: "rs11209026 is genome-wide significantly
                associated with ileal Crohn's disease"
    source_span: {page: 1,
      key_phrase: "rs11209026 (P = 5.05e-9, corrected P = 1.56e-3)"}
  - id: GE-1.A2
    statement: "The Gln allele is protective (OR = 0.26)"
    source_span: {page: 2,
      key_phrase: "glutamine allele appears to protect against
                   development of CD"}
\end{lstlisting}

\paragraph{Candidate extensions.} Curator review identified three candidate
extensions, all provisional under the two-papers promotion rule:

\begin{itemize}
  \item \textbf{CE-DU1, direction of effect:} a first-class dimension
  distinguishing protective from risk-conferring alleles. In this case, the
  protective direction is implicit in an odds ratio of 0.26. The distinction
  may matter for downstream ACMG/AMP-style use.
  \item \textbf{CE-DU3, cross-item synthesis:} a construct for
    higher-order claims that jointly reference several evidence items. Here,
    the relevant claim is the paper's recommendation that blocking the IL-23
    pathway is a rational therapeutic strategy. The issue arises twice, in
    that recommendation and in GE-new-1, and therefore also raises a question
    about the model's translational scope.
  \item \textbf{CE-DU4, conditional-activation scope:} widen the
    activation condition for
    \{\texttt{mode\_of\_\allowbreak inheritance},
    \texttt{mendelian\_\allowbreak segregation}, \texttt{exact\_variant},
    \texttt{subdomain}\} beyond
    \dimval{Human Genetics}~$+$~\dimval{Gene} to also cover
    \dimval{Human Genetics}~$+$~\dimval{Variant}. GE-3, a family-based
    transmission test of a specific variant, illustrates this need. The
    extension would also require a \dimval{Complex} mode-of-inheritance
    value and a family-based subdomain value.
\end{itemize}
A fourth candidate, CE-DU2, proposed a separate dimension for negated
assertions. It was \emph{retracted} on review because an assertion is already
a predicate and can express either absence or presence. This retraction is
the corpus's clearest example of the conservative promotion rule preventing
a premature addition.

\subsection{Case Report CR-IN: Inouye et al.\ 2018 (polygenic risk score)}
\label{sec:supp-ex-inouye}

Inouye and colleagues developed and externally validated a meta-genomic risk
score (metaGRS) for coronary artery disease from 1{,}745{,}180 variants in
482{,}629 UK Biobank participants. The evidential target is a derived,
whole-genome score, and the primary measurements are epidemiological, such
as hazard ratios and C-indices. The paper therefore exposes limitations of
a model whose target types are variant-level.

The annotation reported here is the \texttt{v1} re-annotation produced
under the current protocol, which was refined during the Duerr review. The
original five-item \texttt{v0} draft remains in the repository for
comparison. The \texttt{v1} annotation decomposes the paper into four
claim-coherent items (\cref{tab:cr-inouye}). Each can be represented only
through deliberate placeholder assignments, recorded as forced fits and
candidate extensions rather than left implicit.

\begin{table}[H]
  \centering
  \footnotesize
  \caption[Inouye 2018: evidence-item decomposition]{Inouye 2018
  (\texttt{v1}): four \GeneticEvidence{} items. Each uses the same
  placeholder Target Type and Resolution value, \dimval{Variant}, and is
  flagged against CE-IN1 and CE-IN2. The actual target is a whole-genome
  polygenic score.}
  \label{tab:cr-inouye}
  \begin{tabularx}{\linewidth}{@{}l >{\raggedright\arraybackslash}X p{1.5cm} l@{}}
    \toprule
    ID & Claim & Target$^{\dagger}$ & Cred. \\
    \midrule
    GE-1     & Prediction and quintile stratification (HR 1.71/SD; 4.17-fold) & \dimval{Variant} & High \\
    GE-2     & Independence from six conventional risk factors & \dimval{Variant} & High \\
    GE-3     & Stratification persists among medicated individuals & \dimval{Variant} & High \\
    GE-new-1 & Combining three component scores outperforms each & \dimval{Variant} & Medium \\
    \bottomrule
  \end{tabularx}

  {\footnotesize $^{\dagger}$Placeholder assignment. The actual target is a
  polygenic \dimval{Score} (CE-IN1) at whole-genome-aggregate resolution
  (CE-IN2). Method is recorded at the intermediate
  \dimval{Association Study}/\dimval{Meta-Analysis} level because no leaf
  represents polygenic-score association (flag F-METH). Variant Ascertainment
  is marked not applicable because a score has no
  single ascertainment mode; this supports the CE-IN1 SHACL argument.}
\end{table}

The independence item, GE-2, makes these limitations concrete. Both its
\texttt{target\_type} and \texttt{target\_resolution} are assigned the
placeholder value \dimval{Variant} and flagged against CE-IN1 and CE-IN2.
Variant Ascertainment is marked not applicable because it does not apply
cleanly to a score target. Epidemiological measures are recorded in
\texttt{special\_considerations} and motivate CE-IN4.

The item is shown in abridged form below. Protocol~v0, used for Duerr,
named the fields \texttt{resolution} and \texttt{key\_phrase}; protocol~v1,
used for Inouye, renamed them \texttt{target\_resolution} and
\texttt{phrase} (\cref{sec:supp-protocol-suite}). The corpus retains both
forms, which the YAML-to-RDF transformation normalizes to
\texttt{gem:resolution} and \texttt{gem:phrase}; \texttt{gem-validate}
therefore treats them identically.

\begin{lstlisting}[language=yaml]
id: GE-2
label: "metaGRS predicts incident CAD largely independently
        of six conventional risk factors"
knowledge_domain: [POPULATION_GENETICS]
method: [ASSOCIATION_STUDY]    # intermediate; no leaf fits PGS assoc. (F-METH)
target_type: VARIANT           # forced; CE-IN1 proposes SCORE
target_resolution: VARIANT     # forced; CE-IN2 -> WHOLE_GENOME_AGGREGATE
variant_ascertainment: not_applicable_or_omitted  # CE-IN1 (SHACL); F-VA
phenotype_scale: CLINICAL
credibility: HIGH
# epidemiological measures (C-index, HR) in special_considerations; CE-IN4
assertions:
  - id: GE-2.A1
    statement: "metaGRS has a higher C-index than any single factor"
    source_span: {page: 6,
      phrase: "higher C-index (C = 0.623; 95% CI: 0.615 to 0.630)"}
  - id: GE-2.A3
    statement: "Adding metaGRS to all six factors raised C-index to 0.696"
    source_span: {page: 6,
      phrase: "C-index of 0.696 (95% CI: 0.688 to 0.703)"}
\end{lstlisting}

\paragraph{Candidate extensions.} The Inouye annotation identified six
candidate extensions, the most of any paper in the corpus:

\begin{itemize}
  \item \textbf{CE-IN1, \dimval{Score} target type:} a metaGRS is
  neither a gene nor a variant, but a derived composite predictor. A
  \dimval{Score} value, possibly characterized as polygenic, composite, or
  both, would represent the target directly. Under the current
  \dimval{Variant} placeholder, all four items mark Variant Ascertainment
  as not applicable. The schema accepts this as an explicit escape, but a
  first-class \dimval{Score} target would remove the forced fit.

  \item \textbf{CE-IN2, \dimval{Whole Genome Aggregate} resolution:} the
  score aggregates variants across the genome non-contiguously, which the
  current contiguous or pointwise Resolution values do not capture.

  \item \textbf{CE-IN3, target-composition dimension:} an axis orthogonal
  to Resolution that distinguishes \dimval{Single},
  \dimval{Aggregate}, and \dimval{Composite} targets. This provides a
  complementary framing of the same underlying gap as CE-IN2.

  \item \textbf{CE-IN4, epidemiological measurement targets:} an
  extension of \texttt{measurement\_target}, which is currently oriented
  toward molecular assays, with values such as \dimval{Hazard Ratio},
  \dimval{Odds Ratio}, \dimval{C-Index}, and \dimval{AUROC}. These values
  would be activated for population-genetics evidence.

  \item \textbf{CE-IN5, structured cohort descriptor:} a replacement for
  free-text cohort properties with named fields for size, ancestry
  composition, follow-up, and ascertainment. Such a structure could be
  reused across Duerr, Davis, and future cohort studies.

  \item \textbf{CE-IN6, natural versus derived-artifact target:} a flag,
  or a \dimval{Natural}/\dimval{Derived} dimension, indicating that the
  metaGRS is a human-designed construct rather than a natural kind.
\end{itemize}

\section{Dimension coverage over the annotated corpus}
\label{sec:supp-coverage-note}

This note reports how often each dimension is populated across the 28
\GeneticEvidence{} items in the pilot corpus and how often each conditional
dimension is applicable under its activation rules. The counts are computed
from the released YAML annotations by
\fpath{scripts/compute_coverage.py}. A per-paper report is available in the
repository at
\fpath{annotations/coverage.md}.\footnote{\url{https://github.com/ForomePlatform/genetic-evidence-model/blob/master/annotations/coverage.md}}

\begin{table}[H]
  \centering
  \small
  \caption[Dimension population across 28 \GeneticEvidence{} items]{Dimension
  population across 28 \GeneticEvidence{} items from six papers
  (Duerr \texttt{v0}, Inouye \texttt{v1}), computed by
  \texttt{compute\_coverage.py}. The left block lists always-required
  dimensions; the right block lists conditional dimensions and the number of
  items in which their activation conditions apply.}
  \label{tab:supp-coverage}
  \begin{tabular}{@{}lcc@{\hspace{1em}}lcc@{}}
    \toprule
    \multicolumn{3}{c}{Always-required dimensions} &
    \multicolumn{3}{c}{Conditional dimensions} \\
    Dimension             & Populated & Applicable &
    Dimension             & Populated & Applicable \\
    \midrule
    Knowledge Domain      & 27        & 28         &
    Variant Ascertainment & 14        & 18         \\
    Method                & 28        & 28         &
    Mode of Inheritance   & 0         & 1          \\
    Target Type           & 28        & 28         &
    Mendelian Segregation & 0         & 1          \\
    Resolution            & 28        & 28         &
    Penetrance            & 2         & 4          \\
    Credibility           & 28        & 28         &
    Measurement Target    & 10        & 11         \\
    Phenotype Scale       & 28        & 28         &
    Gene Relation         & 1         & 11         \\
                          &           &            &
    Organism              & 8         & 8          \\
                          &           &            &
    Knockout Type         & 1         & 6          \\
    \bottomrule
  \end{tabular}
\end{table}

The always-required dimensions are populated in all 28 items by
construction, except Knowledge Domain, which is populated in 27. The
exception is a cited anti-p40 antibody trial in Duerr, which concerns
drug-response evidence with no fitting genetic Knowledge Domain.

Among the conditional dimensions, Variant Ascertainment is populated in 14 of
the 18 items for which its activation condition holds. The four Inouye
\texttt{v1} items use \dimval{Variant} as a placeholder Target Type but mark
Variant Ascertainment as not applicable, because a polygenic score has no
single ascertainment mode. This is the CE-IN1 gap. Well-calibrated conditions
such as Organism, populated in 8 of 8 applicable items, and Measurement
Target, populated in 10 of 11, are generally populated whenever their
conditions hold. In contrast, Gene Relation, populated in 1 of 11 applicable
items, and Knockout Type, populated in 1 of 6, expose enumeration and
over-broad-activation gaps that motivate candidate extensions.

Mode of Inheritance and Mendelian Segregation are each populated in 0 of 1
applicable items. The single \dimval{Human Genetics}~$+$~\dimval{Gene} item,
Davis's case--control burden test, addresses both as not applicable. This
indicates that the current activation condition is broader than the context in
which those dimensions are meaningful.

    \section{UMLS crosswalk and illustrative OMOP CDM positioning}
\label{sec:s-umls-omop}

\subsection{UMLS concept crosswalk (generated)}
\label{sec:s-umls-crosswalk}

This crosswalk binds Genetic Evidence Model dimensions to Unified Medical Language System (UMLS) Semantic Types and maps enumerated values to UMLS concepts where an adequate mapping is available. It is produced by the mapping harness (\texttt{gem-umls-crosswalk}), which resolves each term against the UMLS Metathesaurus through the UTS REST API or against a locally indexed licensed copy, and by curator adjudication in the accompanying curation studio (\texttt{gem-mapping-studio}). The machine-readable sources are \texttt{data/umls/umls\_crosswalk.yaml} (the harness output) and \texttt{data/umls/adjudications.yaml} (the curator's decisions). The narrative behind the decisions is kept in the repository's \texttt{data/umls/DECISIONS.md}.

This UMLS crosswalk is distinct from the structural standards crosswalk in \Cref{sec:supp-crosswalk} and does not replace it.

A concept is reported as mapped only when the harness retrieved it and it was confirmed against UMLS. No concept identifier is asserted unless it was returned by a query and confirmed against UMLS.

Each dimension \emph{axis} is bound to a UMLS Semantic Type. The type scopes the search for that dimension's values; a curator may deliberately accept a concept outside the axis and records why. Each dimension \emph{value} is mapped to a Metathesaurus concept. The \emph{relation} column records how the accepted concept relates to the GEM token: \emph{exact}, \emph{close}, \emph{narrower} (the GEM token is more specific than the concept), \emph{broader} (the GEM token is more general), or \emph{related} (a constituent or operator concept anchors a composite token).

An unmapped value is an adjudicated conclusion, not an absence. Relation \emph{none} means that no faithful UMLS mapping was accepted. The verdict is supported by rejected near-miss concepts, each assigned the adequacy criterion it failed, and by the documented search protocol that found no proxy. Where a concept identifier appears in connection with an argued gap, it identifies a retrieved or rejected near miss, not an accepted mapping. The credibility tiers are argued in full in \Cref{sec:supp-credibility}.

\noindent\textbf{Coverage.} The model has 17 dimension axes; 14 are bound to a UMLS Semantic Type, and 3 boolean or free-text dimensions take no type by design. Of the 71 enumerated values, 62 carry an accepted mapping and 9 are argued gaps with a structured rationale (\emph{method}: FAMILY\_BASED, TRANSMISSION\_DISEQUILIBRIUM\_TEST, SEGREGATION\_ANALYSIS; \emph{credibility}: VERY\_HIGH, HIGH, MEDIUM, LOW; \emph{measurement\_target}: EXISTENCE; \emph{gene\_relation}: X\_has\_same\_function\_as\_Y).

{\footnotesize
\begin{longtable}{@{}l l l p{7.2cm}@{}}
\caption{GEM dimensional vocabulary mapped to UMLS concepts (generated by the mapping harness and curator adjudication).}\\
\toprule
GEM value & CUI / TUI & Relation & UMLS concept or semantic type \\
\midrule
\endfirsthead
\toprule
GEM value & CUI / TUI & Relation & UMLS concept or semantic type \\
\midrule
\endhead
\midrule \multicolumn{4}{@{}l}{\textbf{knowledge\_domain}} \\
\quad (axis) & T090 & --- & Occupation or Discipline \emph{(semantic type, tree A2.6)} \\
\quad HUMAN\_GENETICS & C0020124 & exact & Human Genetics \emph{(Biomedical Occupation or Discipline; MSH)} \\
\quad ANIMAL\_GENETICS & C1510895 & exact & Animal Genetics \emph{(Biomedical Occupation or Discipline; NCI)} \\
\quad POPULATION\_GENETICS & C0017404 & exact & Genetics, Population \emph{(Occupation or Discipline; MSH)} \\
\quad COMPARATIVE\_GENOMICS & C4704937 & exact & Comparative Genomics \emph{(Biomedical Occupation or Discipline; MSH)} \\
\quad EPIGENETICS & C1655731 & exact & study of epigenetics \emph{(Biomedical Occupation or Discipline; MTH)} \\
\quad GENE\_FUNCTION & C0872316 & exact & Functional Genomics \emph{(Biomedical Occupation or Discipline; MTH)} \\
\quad MODEL\_ORGANISM & C0599779 & exact & Animal Model \emph{(Animal; MTH)} \\
\midrule \multicolumn{4}{@{}l}{\textbf{method}} \\
\quad (axis) & T062 & --- & Research Activity \emph{(semantic type, tree B1.3.2)} \\
\quad STATISTICAL\_GENETICS & C2717898 & narrower & Biostatistics \emph{(Research Activity; MTH)} \\
\quad ASSOCIATION\_STUDY & C2717878 & exact & Genetic Association Studies \emph{(Molecular Biology Research Technique; MSH)} \\
\quad GWAS & C2350277 & exact & Genome-Wide Association Study \emph{(Molecular Biology Research Technique; MSH)} \\
\quad CANDIDATE\_GENE\_STUDY & C2717881 & exact & Candidate Gene Identification \emph{(Molecular Biology Research Technique; MSH)} \\
\quad FINE\_MAPPING & C1517888 & close & Linkage Disequilibrium Mapping \emph{(Molecular Biology Research Technique; MTH)} \\
\quad FAMILY\_BASED & --- & \emph{none} & \emph{no faithful concept; argued gap (search protocol recorded)} \\
\quad LINKAGE\_ANALYSIS & C0796345 & exact & genetic linkage analysis \emph{(Laboratory Procedure; MTH)} \\
\quad TRANSMISSION\_DISEQUILIBRIUM\_TEST & --- & \emph{none} & \emph{no faithful concept; argued gap (2 rejected near-misses)} \\
\quad SEGREGATION\_ANALYSIS & --- & \emph{none} & \emph{no faithful concept; argued gap (3 rejected near-misses)} \\
\quad META\_ANALYSIS & C0920317 & exact & Meta-Analysis (statistical procedure) \emph{(Research Activity; MTH)} \\
\quad EXPERIMENT & C0678323 & exact & experimentation \emph{(Research Activity; MTH)} \\
\quad IN\_VIVO & C0681829 & exact & in vivo study \emph{(Research Activity; MTH)} \\
\quad IN\_VITRO & C0681828 & exact & in vitro study \emph{(Research Activity; MTH)} \\
\quad IN\_SILICO & C0009609 & exact & Computer simulation \emph{(Machine Activity; MTH)} \\
\quad BIOINFORMATICS\_INFERENCE & C0376528 & narrower & Computational Biology \emph{(Biomedical Occupation or Discipline; MTH)} \\
\quad CLINICAL\_EVIDENCE & C0008972 & exact & Clinical Research \emph{(Research Activity; MTH)} \\
\midrule \multicolumn{4}{@{}l}{\textbf{target\_type}} \\
\quad (axis) & T028 & --- & Gene or Genome \emph{(semantic type, tree A1.2.3.5)} \\
\quad GENE & C0017337 & exact & Genes \emph{(Gene or Genome; MTH)} \\
\quad RELATED\_GENE & C0332281 & related & Associated with \emph{(Qualitative Concept; MTH)} \\
\quad VARIANT & C0042333 & close & Genetic Variation \emph{(Natural Phenomenon or Process; MTH)} \\
\quad SEGMENT & C0678933 & close & Genetic Loci \emph{(Gene or Genome; MTH)} \\
\quad INTERVAL & C1707511 & related & Coordinate on an axis or a grid \emph{(Spatial Concept; MTH)} \\
\quad TRANSCRIPT & C1519595 & close & RNA Transcript \emph{(Nucleic Acid, Nucleoside, or Nucleotide; MTH)} \\
\midrule \multicolumn{4}{@{}l}{\textbf{resolution}} \\
\quad (axis) & T082 & --- & Spatial Concept \emph{(semantic type, tree A2.1.5)} \\
\quad WINDOW & C0678933 & close & Genetic Loci \emph{(Gene or Genome; MTH)} \\
\quad GENE & C1708726 & exact & Gene Locus \emph{(Spatial Concept; MTH)} \\
\quad FUNCTIONAL\_ELEMENT & C1517495 & close & Gene Feature \emph{(Gene or Genome; MTH)} \\
\quad POSITION & C1707511 & exact & Coordinate on an axis or a grid \emph{(Spatial Concept; MTH)} \\
\quad VARIANT & C0042333 & close & Genetic Variation \emph{(Natural Phenomenon or Process; MTH)} \\
\midrule \multicolumn{4}{@{}l}{\textbf{credibility}} \\
\quad (axis) & T080 & --- & Qualitative Concept \emph{(semantic type, tree A2.1.2)} \\
\quad VERY\_HIGH & C0205250 & \emph{none} & \emph{no faithful concept; argued gap (11 rejected near-misses)} \\
\quad HIGH & C0205250 & \emph{none} & \emph{no faithful concept; argued gap (11 rejected near-misses)} \\
\quad MEDIUM & C1561547 & \emph{none} & \emph{no faithful concept; argued gap (11 rejected near-misses)} \\
\quad LOW & C0205251 & \emph{none} & \emph{no faithful concept; argued gap (11 rejected near-misses)} \\
\midrule \multicolumn{4}{@{}l}{\textbf{phenotype\_scale}} \\
\quad (axis) & T071 & --- & Entity \emph{(semantic type, tree A)} \\
\quad MOLECULAR & C0567416 & related & Molecule \emph{(Substance; MTH)} \\
\quad CELLULAR & C0007634 & related & Cells \emph{(Cell; MTH)} \\
\quad HISTOLOGICAL & C0040300 & related & Body tissue \emph{(Tissue; MTH)} \\
\quad ORGANISMAL & C0029235 & related & Organism \emph{(Organism; MTH)} \\
\quad CLINICAL & C0037088 & related & Signs and Symptoms \emph{(Sign or Symptom; MTH)} \\
\midrule \multicolumn{4}{@{}l}{\textbf{variant\_ascertainment}} \\
\quad (axis) & T062 & --- & Research Activity \emph{(semantic type, tree B1.3.2)} \\
\quad OBSERVED\_IN\_CASES & C0007328 & related & Case-Control Studies \emph{(Research Activity; MTH)} \\
\quad OBSERVED\_IN\_CONTROLS & C0007328 & related & Case-Control Studies \emph{(Research Activity; MTH)} \\
\quad FROM\_DATABASE & C0993637 & exact & Published Database \emph{(Intellectual Product; MTH)} \\
\quad SYNTHETIC & C0598279 & exact & synthetic construct \emph{(Nucleic Acid, Nucleoside, or Nucleotide; MTH)} \\
\midrule \multicolumn{4}{@{}l}{\textbf{mode\_of\_inheritance}} \\
\quad (axis) & T045 & --- & Genetic Function \emph{(semantic type, tree B2.2.1.1.4.1)} \\
\quad autosomal\_dominant & C0443147 & exact & Autosomal dominant inheritance \emph{(Genetic Function; MTH)} \\
\quad autosomal\_recessive & C0441748 & exact & Autosomal recessive inheritance \emph{(Genetic Function; SNOMEDCT\_US)} \\
\quad x\_linked\_dominant & C1847879 & exact & X-linked dominant inheritance \emph{(Finding; MTH)} \\
\quad x\_linked\_recessive & C1845977 & exact & X-linked recessive inheritance \emph{(Finding; MTH)} \\
\quad mitochondrial & C0887941 & exact & Mitochondrial Inheritance \emph{(Genetic Function; MTH)} \\
\midrule \multicolumn{4}{@{}l}{\textbf{penetrance}} \\
\quad (axis) & T045 & --- & Genetic Function \emph{(semantic type, tree B2.2.1.1.4.1)} \\
\quad complete & C5826795 & exact & Typified by complete penetrance \emph{(Genetic Function; HPO)} \\
\quad incomplete & C1836598 & exact & Typified by incomplete penetrance \emph{(Finding; MTH)} \\
\quad unknown & C0439673 & close & Unknown \emph{(Qualitative Concept; MTH)} \\
\midrule \multicolumn{4}{@{}l}{\textbf{measurement\_target}} \\
\quad (axis) & T044 & --- & Molecular Function \emph{(semantic type, tree B2.2.1.1.4)} \\
\quad EXISTENCE & C1159832 & \emph{none} & \emph{no faithful concept; argued gap (3 rejected near-misses)} \\
\quad EXPRESSION & C0017262 & exact & Gene Expression \emph{(Genetic Function; MTH)} \\
\quad STABILITY & C2350440 & close & Protein Stability \emph{(Qualitative Concept; MTH)} \\
\quad BINDING & C0033618 & exact & Protein Binding \emph{(Molecular Function; MSH)} \\
\quad LOCALIZATION & C0597704 & close & protein localization activity \emph{(Cell Function; MTH)} \\
\quad ACTIVITY & C1148560 & narrower & molecular\_function \emph{(Molecular Function; GO)} \\
\quad CATALYSIS & C0243102 & exact & enzyme activity \emph{(Molecular Function; MTH)} \\
\midrule \multicolumn{4}{@{}l}{\textbf{gene\_relation}} \\
\quad (axis) & T045 & --- & Genetic Function \emph{(semantic type, tree B2.2.1.1.4.1)} \\
\quad X\_has\_same\_function\_as\_Y & --- & \emph{none} & \emph{no faithful concept; argued gap (search protocol recorded)} \\
\quad X\_regulates\_Y & C0017263 & exact & Gene Expression Regulation \emph{(Genetic Function; MTH)} \\
\quad X\_inhibits\_Y & C2611924 & related & negative regulation of gene expression \emph{(Genetic Function; GO)} \\
\midrule \multicolumn{4}{@{}l}{\textbf{organism}} \\
\quad (axis) & T001 & --- & Organism \emph{(semantic type, tree A1.1)} \\
\midrule \multicolumn{4}{@{}l}{\textbf{knockout\_type}} \\
\quad (axis) & T063 & --- & Molecular Biology Research Technique \emph{(semantic type, tree B1.3.2.1)} \\
\quad CONDITIONAL & C0814041 & exact & conditional gene knockout technology \emph{(Molecular Biology Research Technique; AOD)} \\
\quad UNCONDITIONAL & C0599772 & narrower & Gene Knockout Techniques \emph{(Molecular Biology Research Technique; MTH)} \\
\midrule \multicolumn{4}{@{}l}{\textbf{subdomain}} \\
\quad (axis) & T063 & --- & Molecular Biology Research Technique \emph{(semantic type, tree B1.3.2.1)} \\
\quad GWAS & C2350277 & exact & Genome-Wide Association Study \emph{(Molecular Biology Research Technique; MSH)} \\
\quad Linkage Study & C0796345 & exact & genetic linkage analysis \emph{(Laboratory Procedure; MTH)} \\
\quad WGS-WES Study & C3640076 & broader & Whole Genome Sequencing \emph{(Molecular Biology Research Technique; MSH)} \\
\quad Candidate Gene Study & C2717881 & exact & Candidate Gene Identification \emph{(Molecular Biology Research Technique; MSH)} \\
\midrule \multicolumn{4}{@{}l}{\textbf{mendelian\_segregation}} \\
\quad (axis) & --- & --- & \emph{no semantic type (boolean / free-text dimension)} \\
\midrule \multicolumn{4}{@{}l}{\textbf{genetic\_background\_considered}} \\
\quad (axis) & --- & --- & \emph{no semantic type (boolean / free-text dimension)} \\
\midrule \multicolumn{4}{@{}l}{\textbf{environmental\_factors}} \\
\quad (axis) & --- & --- & \emph{no semantic type (boolean / free-text dimension)} \\
\bottomrule
\end{longtable}
}

\subsection{OMOP CDM positioning (illustrative)}
\label{sec:s-omop-positioning}

The following worked examples position GEM dimensions relative to the OHDSI OMOP Common Data Model (CDM) and its standardized vocabularies. This is an illustrative, hand-authored positioning exercise, not a full crosswalk. OMOP CDM is built around observational, patient-level clinical data, including conditions, drugs, measurements, and observations. GEM describes basic-science genetic evidence about gene--phenotype relationships, much of which has no native patient-level representation in the CDM.

{\footnotesize
\begin{longtable}{@{}p{2.9cm} p{2.6cm} p{2.1cm} p{6.0cm}@{}}
\caption{Illustrative positioning of GEM dimensions against the OMOP CDM (hand-authored, not exhaustive).}\\
\toprule
GEM dimension & OMOP domain / table & Standard vocab. & Alignment and gap \\
\midrule
\endfirsthead
\toprule
GEM dimension & OMOP domain / table & Standard vocab. & Alignment and gap \\
\midrule
\endhead
phenotype\_scale (CLINICAL) & Condition\newline condition\_occurrence & SNOMED CT & A clinical-scale phenotype aligns with an OMOP Condition concept (SNOMED CT), the CDM's native representation of a clinical finding. \emph{Gap:} Molecular, cellular, histological and organismal phenotype scales have no Condition-domain representation; OMOP has no phenotype-scale axis. \\
\addlinespace
measurement\_target (EXPRESSION, ACTIVITY, ...) & Measurement\newline measurement & LOINC & A quantitative molecular readout is conceptually a Measurement; LOINC covers some assay readouts (e.g. enzyme activity panels). \emph{Gap:} Basic-science functional readouts (binding, localization, stability of a specific gene product in an experimental system) are rarely represented as standard LOINC Measurement concepts. \\
\addlinespace
target\_type (GENE) / resolution (GENE) & Observation / (Genomic CDM extension)\newline observation (or OMOP Genomic CDM) & HGNC (via the OMOP Genomic vocabulary effort) & A gene identity aligns with HGNC; the OHDSI Genomic CDM extension is the locus where gene- and variant-level data are being standardized. \emph{Gap:} The base OMOP CDM has no first-class gene or variant entity; this lives only in the evolving Genomic CDM extension, which is not yet a stable standard. \\
\addlinespace
organism & (none)\newline (none) & (none in CDM; NCBITaxon outside it) & No alignment. OMOP CDM is patient-centric (Homo sapiens implied) and has no concept of a model organism. \emph{Gap:} Model-organism evidence is entirely out of scope for the CDM; GEM's organism dimension maps to NCBITaxon (\Cref{sec:supp-crosswalk}), not OMOP. \\
\addlinespace
method (GWAS, ASSOCIATION\_STUDY, ...) & (metadata / study-level, not CDM clinical domains)\newline (none; study design is not a CDM patient-data domain) & (none; partial coverage in MeSH/OBI outside OMOP) & No direct alignment. Study design / evidence-generation method is not a patient-level CDM domain. \emph{Gap:} The CDM records observations, not how upstream genetic evidence was generated; GEM's method hierarchy has no OMOP counterpart. \\
\addlinespace
\bottomrule
\end{longtable}
}

    \section{Credibility tiers: definitions, the UMLS finding, and external alignment}
\label{sec:supp-credibility}

\subsection{What the tiers mean}
\label{sec:supp-credibility-defs}

The overall credibility rating is the model's only subjective core
dimension. Its four levels are defined by \emph{defeasibility}: what
would lead a curator to stop accepting an evidence item. They are not
defined by study design or by the certainty expressed by the authors:

\begin{description}
  \item[\dimval{VERY\_HIGH}] accepted by default even in the face of direct
    contradiction; reconsidered only in light of a stronger opposing
    argument;
  \item[\dimval{HIGH}] accepted unless directly contradicted by evidence of
    comparable strength; such a contradiction calls the item into question;
  \item[\dimval{MEDIUM}] not accepted on its own; accepted when corroborated
    by an independent source or an orthogonal type of evidence;
  \item[\dimval{LOW}] not relied upon; recorded pending independent
    replication.
\end{description}

The distinction between \dimval{HIGH} and \dimval{VERY\_HIGH} is
deliberate. Evidence of comparable strength is sufficient to call a
\dimval{HIGH} item into question, whereas a \dimval{VERY\_HIGH} item
remains the default position unless the opposing case is stronger.

This definition is a belief-revision policy. It therefore corresponds to
SEPIO's confidence attribute: the degree of belief that an asserted
proposition is true~\cite{sepio}. The credibility facets, including cohort
size, replication, multiple-testing control, ancestry control, and
ascertainment, inform that judgment but remain separate.

This separation is inspired by GRADE's general practice of assessing the
factors that affect certainty separately~\cite{grade2008}. GEM does not
adopt the GRADE framework or its rating rules, which were developed for
bodies of clinical evidence, particularly evidence from randomized and
observational studies. The GEM facets instead support credibility judgments
about individual evidence items from basic and pre-clinical research.

The scale intensifies at the top, as ACMG evidence strength
(\emph{supporting} to \emph{very strong})~\cite{acmg2015} and ClinGen
gene--disease validity (\emph{limited} to \emph{definitive})~\cite{clingen2018}
do. GRADE certainty instead extends downward to \emph{very low}, because it
rates a body of evidence by downgrading from a starting level determined by
study design. The GEM and GRADE scales therefore address related but
non-equivalent constructs; their alignment in \cref{tab:cred-alignment} is
deliberately non-positional.

\subsection{Why the tiers have no UMLS proxy}
\label{sec:supp-credibility-umls}

The four credibility tiers have no adequate UMLS proxy. This is a curation
conclusion, not merely a missing mapping: the tiers were tested in a
pre-registered sweep (\texttt{data/umls/sweeps/credibility.yaml}).

The adequacy criteria were fixed before the search. A candidate concept had
to (A) denote a degree of epistemic warrant rather than the magnitude of a
measured property; (B) represent a scale point rather than the scale
dimension itself; (C) belong to a set that is disjoint and ordered in the
same way as the GEM scale; and (D) carry that meaning in the relevant domain
rather than as a homonym.

The sweep used a local index of UMLS 2026AA and comprised 174 queries that
returned 497 distinct concepts. The results were replicated through the UTS
REST API, which returned 505 concepts. The search had three passes:
whole-string searches for intensified and degree qualifiers; searches for
the names of evidence-grading vocabularies; and searches for the same terms
within the UMLS semantic types most likely to contain scale values:
Qualitative Concept, Quantitative Concept, Idea or Concept, Intellectual
Product, and Finding. Each
retrieved concept was assigned the criterion it failed
(\cref{tab:cred-families}); the full generated table is available in the
results file.

\begin{table}[H]
  \centering
  \small
  \caption[Credibility sweep: rejected concept families]{Concepts retrieved by
  the credibility sweep (UMLS 2026AA), grouped by the adequacy criterion they
  failed. Criterion codes are defined in the text.}
  \label{tab:cred-families}
  \begin{tabularx}{\linewidth}{@{}>{\raggedright\arraybackslash}X r c >{\raggedright\arraybackslash}X@{}}
    \toprule
    Concept family & $n$ & Criterion failed & Examples \\
    \midrule
        “grade” (tumour, school, or product) & 202 & D & Astrocytoma, low grade (C1314694) \\
        LOINC/PROMIS self-efficacy items & 191 & D & Current level of confidence I can drive a car (C5142440) \\
        SNOMED degree qualifier values & 21 & A & Very high (C0442804); Extremely high (C5787630) \\
        clinical “evidence of” expressions (presence of a sign) & 20 & D & Blood alcohol level of less than 20 mg/100 ml (C0481372); Blood alcohol level of 20--39 mg/100 ml (C0481373) \\
        qualifiers denoting the degree of a measured property & 12 & A & Observation Value - High (C1561957); Message Waiting Priority - High (C1561958) \\
        scores, probabilities, risk categories & 8 & A & IPSS-R Risk Category Very High (C5202916); IPSS Risk Category High (C4522209) \\
        patient-perceived credibility of information & 7 & D & Establish teacher credibility, as appropriate (C0511083); credibility (C0870373) \\
        NCI/PDQ Level of Evidence I--IV & 6 & A & Level of Evidence (C0393009); Level of Evidence I (C2986612) \\
        metabolite-identification confidence levels & 6 & D & Level 1 Metabolite Identification Confidence (C5551417); Level 3 Metabolite Identification Confidence (C5551419) \\
        SNOMED diagnostic-certainty modality & 5 & C & Definite (C0439544); Probable diagnosis (C0332148) \\
        NCI confidence answer set & 4 & D & Very High Confidence (C4331500); High Confidence (C4330261) \\
        psychological self-confidence & 2 & D & Euphoric mood (C0235146); Confidence level (self-esteem) (C0518578) \\
        lexical collisions (“media”, “lower”) & 2 & D & Communications Media (C0009458); Culture Media (C0010454) \\
        degree of a measured property & 2 & A & A Medium Amount of Time (C4522282); A Medium Amount (C4522283) \\
        NCI adverse-event causality ladder & 2 & D & Definitely Related to Intervention (C1704787); Possibly Related to Intervention (C1705910) \\
        LOINC molecular-pathology “level of evidence” & 2 & D & Level of evidence:Find:Pt:Bld/Tiss:Nom (C4760312) \\
        diagnostic-certainty rating items & 2 & D & UHDRS 1999 Version - Diagnosis Confidence Level (C4744064); Level of Diagnostic Certainty (C5909371) \\
        TNM certainty factor & 2 & D & Degree of certainty of TNM classification (C0456878); Qualifier for TNM degree of certainty (C1302640) \\
        statistical confidence measures & 1 & A & Level of Confidence (statistical) (C5702556) \\
    \bottomrule
  \end{tabularx}
\end{table}

Four families merit explicit discussion because they are the closest
constructs retrieved from UMLS. First, the SNOMED qualifier values
\emph{High}~(C0205250, SNOMED 75540009), \emph{Very high}~(C0442804), and
\emph{Low}~(C0205251), together with related terms, occur under
\emph{Increased} and \emph{Decreased}. The source terminology therefore
treats them as degrees of a measured property, not as degrees of epistemic
warrant (A). Second, NCI Thesaurus \emph{Level of Evidence I--IV} is a
study-design hierarchy. Mapping GEM credibility to it would collapse the
model's separate \emph{method} and credibility dimensions (A, and C for the
scale). Third, the NCI \emph{Very High / High / Moderate / Low Confidence}
set has the same four-level, top-intensified form as the GEM scale, but it is
a \emph{Clinical or Research Assessment Answer}: a questionnaire response
that records a respondent's confidence in performing an activity (D).

The SNOMED diagnostic-certainty terms \emph{definite}, \emph{probable}, and
\emph{possible} are the closest epistemic candidates. They express the
likelihood that a proposition is true, but they form a three-level scale with
no intensified top category, apply in their source terminologies to
diagnostic assertions, and assess likelihood rather than defeasibility (C,
D). The same limitation applies to the TNM certainty factor, the LOINC
molecular-pathology \emph{level of evidence}, which is an
AMP/ASCO/CAP actionability tier for variant reports, and
metabolite-identification confidence levels. These are ordered epistemic scales, but
each is defined for a different object.

Lexical post-coordination does not supply the missing identifiers. Combining
a qualifier concept with an intensifier from the SPECIALIST Lexicon
(\texttt{LRMOD}) would add only a lexical modifier. The lexicon's
\emph{intensifier} class is a syntactic category that includes
\emph{very}, \emph{slightly}, \emph{about}, and \emph{ago}; it encodes
neither magnitude nor direction. Such an identifier therefore adds no
semantic content beyond the corresponding text string. This is a conceptual
argument about the role of identifiers, not an empirical finding.

\subsection{External alignment}
\label{sec:supp-credibility-align}

The following table gives a conceptual, non-positional alignment between the
GEM tiers and selected external scales. It does not define equivalences or a
common numerical conversion. In particular, the FHIR values are export codes:
\dimval{VERY\_HIGH} must be represented as \texttt{high}, and that export is
therefore lossy.

\begin{table}[H]
  \centering
  \small
  \caption[Credibility: external alignment]{Conceptual, non-positional
  alignment of the GEM credibility tiers with selected external scales.
  ``---'' indicates that no counterpart is proposed. The FHIR column gives the
  lossy export used when a \texttt{certainty-rating} code is
  required~\cite{fhir_evidence}. CIO = Confidence Information
  Ontology~\cite{cio2015}.}
  \label{tab:cred-alignment}
  \begin{tabularx}{\linewidth}{@{}l >{\raggedright\arraybackslash}X >{\raggedright\arraybackslash}X >{\raggedright\arraybackslash}X >{\raggedright\arraybackslash}X >{\raggedright\arraybackslash}X@{}}
    \toprule
    GEM & GRADE certainty~\cite{grade2008} & FHIR \texttt{certainty-rating} & CIO confidence level~\cite{cio2015} & ACMG evidence strength~\cite{acmg2015} & ClinGen validity~\cite{clingen2018} \\
    \midrule
    \dimval{VERY\_HIGH} & --- (GRADE caps at \emph{high}) & \texttt{high} (lossy export) & high (partial) & very
    strong & definitive \\
    \dimval{HIGH}       & high     & \texttt{high}     & high   & strong     & strong \\
    \dimval{MEDIUM}     & moderate & \texttt{moderate} & medium & moderate   & moderate \\
    \dimval{LOW}        & low      & \texttt{low}      & low (``absence of trust'', closer to none) & supporting &
    limited \\
    ---                 & very low & \texttt{very-low} & ---    & ---        & --- \\
    \bottomrule
  \end{tabularx}
\end{table}

These alignments are part of the GEM model, not of the UMLS crosswalk. The
crosswalk records UMLS mappings, or a considered conclusion that no adequate
mapping exists.

    \clearpage
    \part*{Supplementary Tables}
    \addcontentsline{toc}{part}{Supplementary Tables}
    
\begin{table}[H]
    \centering
    \caption[The six annotated publications]{The six annotated publications and the epistemic shape each
    exercises. Manual annotations form the curator-authored reference set; AI-drafted
    annotations are evaluated against curator review. GE abbreviates
    \GeneticEvidence{}; the ``GE items'' column counts how many evidence
    items each paper yields under our claim-level decomposition.}
    \label{tab:corpus}
    \small
    \begin{tabularx}{\linewidth}{@{}lllllX@{}}
\toprule
Paper & Source & GE items & Credibility & Role & Notes \\
\midrule
Jossin 2017 (Llgl1)   & manual       & 6 & Very high & molecular mechanism           & multi-scale phenotype; motivates \emph{phenotype\_scale} \\
Davis 2011 (TTC21B)   & manual       & 6 & High      & breadth exemplar              & motivates \emph{variant\_ascertainment} \\
Nelson 1992 (CD18)    & manual       & 3 & Medium    & classical molecular genetics  & single-gene functional; one candidate (CE-N1) resolved by external reference \\
Gupta 2015 (ATP6AP2)  & manual       & 2 & Low       & low-credibility edge case     & schema fits cleanly \\
Duerr 2006 (IL23R)    & AI-drafted   & 7 & High      & clean GWAS exemplar           & featured in the main text (\texttt{v0}) \\
Inouye 2018 (metaGRS) & AI-drafted   & 4 & High/Med  & model-extension stress test   & featured in the main text (\texttt{v1}) \\
\bottomrule
    \end{tabularx}
\end{table}

\begin{table}[H]
    \centering
    \caption[Credibility facets]{Credibility facets in the \GeneticEvidence{} model.
    Credibility records methodological soundness only; each facet is
    rated separately rather than collapsed to one score, and the facets
    are registered as genetics-specific subcomponents anchored on SEPIO
    confidence. The overall rating's levels are
    analogous in spirit to ClinGen gene-disease validity
    classifications~\cite{clingen2018}, which however aggregate per
    gene-disease relationship rather than per assertion. Effect
    magnitude is not a credibility facet: it is a measured value carried
    on the \GeneticEvidenceVariable{}. Phenotype fit to a clinical
    question is applicability, captured by Relevance and Specificity.}
    \label{tab:credibility-facets}
    \small
    \begin{tabularx}{\linewidth}{@{}lX@{}}
\toprule
Facet & Aspect of methodological soundness recorded \\
\midrule
\texttt{sample\_size}            & Size of cohort, pedigree, or sample (statistical power) \\
\texttt{ascertainment}           & Case/control makeup and selection of the studied sample \\
\texttt{multiple\_testing}       & Whether and how multiplicity was controlled \\
\texttt{stratification\_control} & Ancestry restriction or matching, and residual-confounding control \\
\texttt{replication}             & Independent replication or internal resampling \\
\bottomrule
    \end{tabularx}
\end{table}

\begin{table}[H]
    \centering
    \caption[FHIR field-level alignment]{Field-level alignment of the core model to FHIR Evidence R5. The certainty rows expand the Credibility entry of the class-level alignment table in the main paper.}
    \label{tab:fhir-fields}
    \small
    \begin{tabularx}{\linewidth}{@{}llX@{}}
\toprule
This work & FHIR Evidence R5 & Note \\
\midrule
\ScientificEvidence{} item        & \texttt{Evidence}                              & One evidence concept per claim \\
dimensions                         & \texttt{Evidence.variableDefinition} (roles)   & PICO roles replaced by genetic dimensions \\
\EvidenceVariable{} definition     & \texttt{EvidenceVariable.characteristic}       & What the variable is and how measured \\
\EvidenceVariable{} value          & \texttt{Evidence.statistic}                    & The measured quantity itself \\
Credibility overall rating         & \texttt{Evidence.certainty.rating}             & Coded ordinal rating anchored on SEPIO confidence \\
Credibility facet                  & \texttt{Evidence.certainty.\-certaintySubcomponent.type} & Extension to the subcomponent-type value set \\
facet rating (where graded)        & \texttt{...certaintySubcomponent.rating}       & Per-domain rating \\
\texttt{source\_span}              & \texttt{Evidence.note} / \texttt{Citation}     & Page plus key phrase for audit \\
\bottomrule
    \end{tabularx}
\end{table}

    \bibliographystyle{plainnat}
    \bibliography{references}